\documentclass[a4paper,fleqn,usenatbib]{mnras}

\usepackage{newtxtext,newtxmath}
\usepackage[T1]{fontenc}

\usepackage[english]{babel}
\usepackage{graphicx}
\usepackage{fleqn}
\usepackage{amssymb}
\usepackage{amsmath}
\usepackage{booktabs}
\usepackage{hyperref}
\usepackage{comment}
\usepackage{enumitem}
\setlist[enumerate]{
  labelsep=8pt,
  labelindent=0.\parindent,
 itemindent=0pt,
  leftmargin=*,
}

\usepackage{color, colortbl}

\newcommand*{\mycdot}{\kern-.2em\cdot\kern-.2em}
\renewcommand{\S}{Section}
\newcommand{\F}{Fig.}

\newcommand{\ve}[1]{\boldsymbol{#1}}
\newcommand{\unit}[1]{\hat{\boldsymbol{#1}}}

\newcommand{\msun}{\mathrm{M}_\odot}

\newcommand{\au}{\,\textsc{au}}

\newcommand{\etz}{\ve{\mathcal{E}}_{\scriptscriptstyle \theta,0}}
\newcommand{\jtz}{\ve{\mathcal{J}}_{\scriptscriptstyle \theta,0}}

\newcommand{\ecl}{\ve{\mathcal{E}}^{\scriptscriptstyle c}_{\scriptscriptstyle l}}
\newcommand{\esl}{\ve{\mathcal{E}}^{\scriptscriptstyle s}_{\scriptscriptstyle l}}
\newcommand{\jcl}{\ve{\mathcal{J}}^{\scriptscriptstyle c}_{\scriptscriptstyle l}}
\newcommand{\jsl}{\ve{\mathcal{J}}^{\scriptscriptstyle s}_{\scriptscriptstyle l}}

\newcommand{\etcl}{\ve{\mathcal{E}}^{\scriptscriptstyle c}_{\scriptscriptstyle \theta, l}}
\newcommand{\etsl}{\ve{\mathcal{E}}^{\scriptscriptstyle s}_{\scriptscriptstyle \theta, l}}
\newcommand{\jtcl}{\ve{\mathcal{J}}^{\scriptscriptstyle c}_{\scriptscriptstyle \theta, l}}
\newcommand{\jtsl}{\ve{\mathcal{J}}^{\scriptscriptstyle s}_{\scriptscriptstyle \theta, l}}

\newcommand{\fe}{\ve{f}_{\scriptscriptstyle e}}
\newcommand{\fj}{\ve{f}_{\scriptscriptstyle \j}}

\renewcommand{\ge}{\ve{g}_{\scriptscriptstyle e}}
\newcommand{\gj}{\ve{g}_{\scriptscriptstyle \j}}

\newcommand{\gei}{\ve{g}_{\scriptscriptstyle e}^{\mathrm{I}}}
\newcommand{\geii}{\ve{g}_{\scriptscriptstyle e}^{\mathrm{II}}}
\newcommand{\gji}{\ve{g}_{\scriptscriptstyle \j}^{\mathrm{I}}}
\newcommand{\gjii}{\ve{g}_{\scriptscriptstyle \j}^{\mathrm{II}}}

\renewcommand{\j}{\jmath}

\newcommand{\epssa}{\epsilon_\mathrm{SA}}
\newcommand{\epssade}{\epsilon_{\mathrm{SA},\Delta e=0}}
\newcommand{\epssaplat}{\epsilon_{\mathrm{SA},\mathrm{plateau}}}

\newcommand{\epsoct}{\epsilon_\mathrm{oct}}

\newcommand{\eper}{E}
\newcommand{\mper}{M}
\newcommand{\qper}{Q}

\newcommand{\FO}{\mathrm{\scriptscriptstyle FO}}

\newcommand{\SO}{\mathrm{\scriptscriptstyle SO}}

\newcommand{\oPN}{ {\mathrm{1PN}} }
\newcommand{\tPN}{ {\mathrm{2.5PN}} }
\newcommand{\bin}{ {\mathrm{bin}}}
\newcommand{\rg}{ r_{\mathrm{g}}}
\newcommand{\TB}{ {\mathrm{TB}} }

\newcommand{\Leper}{L}

\newcommand{\lmax}{l_\mathrm{max}}

\pubyear{2019}

\usepackage{listings}

\usepackage{color}

\definecolor{dkgreen}{rgb}{0,0.6,0}
\definecolor{gray}{rgb}{0.5,0.5,0.5}
\definecolor{mauve}{rgb}{0.58,0,0.82}

\allowdisplaybreaks

\begin{document}
\onecolumn
\title[Secular encounters]{Analytic computation of the secular effects of encounters on a binary: features arising from second-order perturbation theory}

\author[Hamers \& Samsing]{Adrian S. Hamers$^{1}$\thanks{E-mail: hamers@ias.edu} and Johan Samsing$^{2}$\thanks{E-mail: jsamsing@gmail.com} \\
$^{1}$Institute for Advanced Study, School of Natural Sciences, Einstein Drive, Princeton, NJ 08540, USA \\
$^{2}$Department of Astrophysical Sciences, Princeton University, Peyton Hall, 4 Ivy Lane, Princeton, NJ 08544, USA}
\date{Accepted 2019 June 9. Received 2019 June 4; in original form 2019 April 24}

\label{firstpage}
\pagerange{\pageref{firstpage}--\pageref{lastpage}}
\maketitle

\begin{abstract}  % 214 words
Binary-single interactions play a crucial role in the evolution of dense stellar systems such as globular clusters. In addition, they are believed to drive black hole (BH) binary mergers in these systems. A subset of binary-single interactions are secular encounters, for which the third body approaches the binary on a relatively wide orbit, and such that it is justified to average the equations of motion over the binary's orbital phase. Previous works used first-order perturbation theory to compute the effects of such secular encounters on the binary. However, this approach can break down for highly eccentric binaries, which are important for BH binary mergers and gravitational wave sources. Here, we present an analytic computation using second-order perturbation techniques, valid to the quadrupole-order approximation. In our calculation, we take into account the instantaneous back-reaction of the binary to the third body, and compute corrections to previous first-order results. Using singly-averaged and direct 3-body integrations, we demonstrate the validity of our expressions. In particular, we show that the eccentricity change for highly eccentric binaries can reach a plateau, associated with a large inclination change, and can even reverse sign. These effects are not captured by previous first-order results. We provide a simple script to conveniently evaluate our analytic expressions, including routines for numerical integration and verification. 
\end{abstract}


\section{Introduction}
\label{sect:introduction}
There exists a large variety of astrophysical settings in which a binary system is perturbed by a passing object in a wider orbit. In the planetary context, stellar flybys can drive wide binaries to high eccentricities, leading to strong tidal interactions and/or stellar collisions \citep{2014ApJ...782...60K}, or destabilizing planetary systems \citep{2013Natur.493..381K}. In the Solar system, stellar encounters play an important role in disturbing the Oort cloud, thus transporting comets into the inner Solar system (\citealt{1950BAN....11...91O}, and, e.g, \citealt{1987Icar...70..269H,1987AJ.....94.1330D,2002A&A...396..283D,2002AN....323...37S,2011Icar..214..334F,2015AJ....150...26H}). Such perturbations may also be responsible for triggering white dwarf pollution in extra-solar systems (e.g., \citealt{2014MNRAS.445.4175V}). In the stellar context, flybys may be important for producing low-mass X-ray binaries \citep{2016MNRAS.458.4188M}, or gravitational wave sources \citep{2019arXiv190201864M}.

In dense stellar systems, flybys in which the binary remains bound after the encounter without exchanges can be considered a sub-type of more general binary-single interactions  (e.g., \citealt{1983ApJ...268..319H,1983ApJ...268..342H,1991MNRAS.250..555H}). Such interactions are key to driving the late-stage evolution of dense stellar systems such as globular clusters (e.g., \citealt{1987degc.book.....S,2008gady.book.....B}). 

If the perturber passes the binary with a periapsis distance that is significantly larger than the binary semimajor axis, then the encounter is typically `secular', i.e., the motion of the perturber is much slower than the orbital motion of the binary. In this case, it is appropriate to expand the Hamiltonian in terms of the ratio of the binary separation to the separation of the perturber to the binary center of mass, and to average the Hamiltonian over the binary motion. These procedures result in the simpler `singly-averaged' (SA) equations of motion, which are computationally less intensive to solve compared to direct 3-body integrations. However, it is still necessary to numerically solve a set of first-order ordinary differential equations (ODEs) in order to obtain the new binary properties after the perturber's passage. 

It is also possible to apply first-order perturbation techniques, i.e., to analytically integrate the SA equations of motion assuming the perturber moves on a hyperbolic or parabolic trajectory, and ignoring changes of the binary's orbital elements during the perturber's passage. This approach was adopted in previous works (e.g., \citealt{1975MNRAS.173..729H,1996MNRAS.282.1064H,2018MNRAS.476.4139H}). In particular, the equations derived by \citet{1996MNRAS.282.1064H} are commonly used to take into account the effects of (distant) encounters on a binary in Monte Carlo-style computations (e.g., \citealt{1995ApJ...445L.133R,2009ApJ...697..458S,2019ApJ...872..165G}). 

However, there are situations in which the first-order perturbation approach can break down. In \citet{1996MNRAS.282.1064H}, it was discussed that their method no longer applies if the eccentricity change is of the order of the initial eccentricity, i.e., $\Delta e \sim e$. However, as we will show here, when the binary is already highly eccentric, then the first-order approach can break down even if $\Delta e$ is small, in particular, if $\Delta e \sim 1-e$. The latter can occur in highly eccentricity binaries perturbed by distant (i.e., weak) encounters, which can drive relatively large changes in the binary's orbital angular momentum. A binary can be driven to high eccentricities as a result of strong (non-secular) encounters in dense stellar systems. The breakdown of the first-order method originates from the fact that the binary's elements can change significantly during the passage of the perturber. Therefore, it is not justified to assume that these elements are constant when integrating the equations of motion. 

In this paper, we present an analytic computation of the change of the binary's properties in the secular regime using second-order perturbation theory. In short, a Fourier expansion is used to describe the binary's instantaneous response to the perturber, and this response is substituted back into the equations of motion in order to obtain correction terms to the first-order result. This approach is similar to that by \citet{2016MNRAS.458.3060L} applied to hierarchical triples (i.e., with a bound third object instead of an unbound one). In triples, similar interactions on the timescale of the third body can modulate the long-term secular interactions (e.g., \citealt{1936MNRAS..97...56B,2004AJ....128.2518C,2005MNRAS.358.1361I,2012arXiv1211.4584K,2012ApJ...757...27A,2013PhRvL.111f1106S,2014MNRAS.439.1079A,2014MNRAS.438..573B,2018MNRAS.481.4602L,2018MNRAS.481.4907G}). The correction terms give rise to behavior such as orbital flips and sign reversals in the eccentricity change that are not captured by first-order theory. 

In an accompanying paper \citep{2019arXiv190607189S}, we investigate the astrophysical implications of secular encounters on highly eccentric binaries, for which the first-order perturbation approach can break down. In particular, the properties of black hole (BH) binary mergers and their subsequent gravitational wave (GW) signals are discussed.

This paper is structured as follows. We give an overview of previous results in \S~\ref{sect:prev}. In \S~\ref{sect:cor}, we apply second-order perturbation techniques, and derive analytic expressions for the changes of the binary's properties in response to secular encounters. These results, which are summarized (for the case of parabolic encounters) in equation~(\ref{eq:Delta_e_CDA_par}), can be interpreted as additions to the work of \citet{1996MNRAS.282.1064H}. In \S~\ref{sect:ex}, we provide a simple \textsc{Python} script to conveniently evaluate our analytic expressions, illustrate the regime in which the first-order approximation breaks down, and demonstrate the correctness of our second-order results using numerically-obtained SA solutions, and direct 3-body integrations. We discuss our results in \S~\ref{sect:discussion}, and conclude in \S~\ref{sect:conclusions}.

\section{Previous results}
\label{sect:prev}
\subsection{Preliminaries}
\label{sect:prev:pre}

\begin{figure}
\center
\includegraphics[scale = 0.65, trim = 0mm 20mm 0mm 10mm]{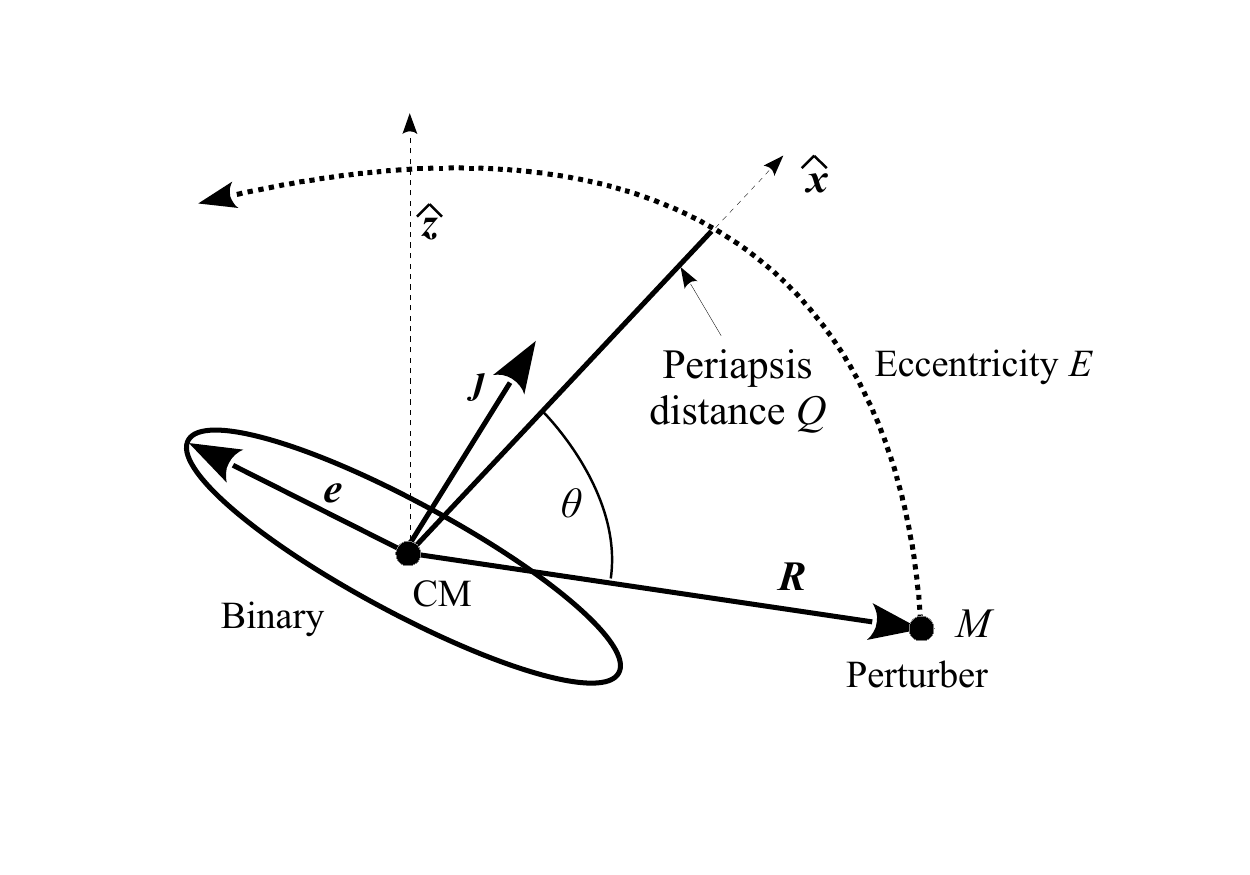}
\caption{Sketch of the configuration. A binary (center of mass `CM')  is perturbed by a passing object (mass $\mper$) on a parabolic or hyperbolic orbit (eccentricity $\eper$) with periapsis distance $Q$ to the binary center of mass ($Q>a$, where $a$ is the binary's semimajor axis). }
\label{fig:sketch}
\end{figure}

Consider a binary with semimajor axis $a$ and eccentricity $e$; the component masses are denoted as $m_1$ and $m_2$, with the total binary mass $m\equiv m_1+m_2$. The instantaneous eccentricity or Laplace-Runge-Lenz vector $\ve{e}$ is given by $\ve{e} = [1/(Gm)] \, \dot{\ve{r}}\times (\ve{r}\times \dot{\ve{r}}) - \unit{r}$, where $\ve{r}$ is the relative separation between the components, dots denote derivatives with respect to time, and hats denote unit vectors. The normalized angular-momentum vector is $\ve{\j}=\ve{r} \times \dot{\ve{r}}$, with magnitude $\j=\sqrt{1-e^2}$. Apart from the orbital phase, the binary state is completely described in terms of $\ve{e}$ and $\ve{\j}$ (note that the six degrees of freedom in these two vectors reduce to effectively four, since $e^2+\j^2=1$, and $\ve{e}\cdot\ve{\j}=0$ -- together with the orbital phase, the binary's state is described by five quantities). 

We assume that the binary is perturbed by a third body with mass $\mper$, passing by on a parabolic or hyperbolic orbit with periapsis distance $\qper>a$, and separation $\ve{R}$ relative to the binary center of mass. In the case of a parabolic (hyperbolic) orbit, the perturber orbit's eccentricity is $\eper=1$ ($\eper>1$). Without loss of generality, we set the angular-momentum vector of the perturber's orbit to be aligned with the $z$-axis, and the eccentricity vector along the $x$-axis (i.e., the periapsis of the perturber's orbit points along the $x$-axis). A sketch of the configuration is shown in \F~\ref{fig:sketch}.

We also assume that the perturber moves sufficiently slowly such that the `secular' approximation applies, i.e., the mean motion of the binary is much faster than the perturber's angular speed at periapsis. 
This condition can be written as (e.g., \citealt{2018MNRAS.476.4139H,2018ApJ...863...68L})
\begin{align}
\label{eq:R_def}
\mathcal{R} = \left [ \left ( 1 + \frac{\mper}{m} \right ) \left (\frac{a}{\qper} \right )^3 \left ( 1 + \eper \right ) \right ]^{1/2} \ll 1.
\end{align}
This condition typically implies that the binary is `hard', i.e., its orbital energy is much larger than the typical kinetic energy of perturbers \citep{1975MNRAS.173..729H}. More quantitatively, the `hardness' of a binary in a cluster with velocity dispersion $\sigma$ can be defined as $h \equiv Gm/(a\sigma^2)$, i.e., the ratio between the (absolute value of the) specific binding energy, and the typical perturber kinetic energy. Writing $\eper = 1 + Q\sigma^2/[G(\mper+m)]$, this implies that $\mathcal{R}$ and $h$ are related according to
\begin{align}
\mathcal{R} = \left [ \left ( 1 + \frac{\mper}{m} \right ) \left (\frac{a}{\qper} \right )^3 \left ( 2 + \frac{Q}{a} \frac{m}{m+\mper} \frac{1}{h} \right ) \right ]^{1/2}.
\end{align}
Therefore, if the binary is hard, $h\gg1$, and
\begin{align}
\mathcal{R} \approx \left [2 \left ( 1 + \frac{\mper}{m} \right ) \left (\frac{a}{\qper} \right )^3 \right ]^{1/2}.
\end{align}
Unless $Q\sim a$ and/or $\mper\gg m$, this implies that $\mathcal{R}\ll1$, i.e., a secular encounter.

Some of the results below are presented in terms of the binary's orbital elements $(i,\omega,\Omega)$, in addition to the orbital vectors $\ve{e}$ and $\ve{\j}$. We define the inclination $i$, argument of periapsis $\omega$, and longitude of the ascending node $\Omega$ in the usual way, i.e., the relation between the orbital angles and orbital vectors is
\begin{subequations}
\begin{align}
\unit{e} &= (\cos \Omega \cos\omega - \sin\Omega\sin\omega\cos i)\,  \unit{x} + (\sin \Omega \cos\omega + \cos\Omega\sin\omega\cos i) \,\unit{y} + \sin\omega \sin i\, \unit{z}; \\
\unit{\j} &= \sin\Omega \sin i \, \unit{x} - \cos \Omega \sin i \, \unit{y} + \cos i \, \unit{z}.
\end{align}
\end{subequations}
We emphasize that our definition of the orbital elements differs from that of \citet{1996MNRAS.282.1064H}. The latter authors chose the binary orientation to be fixed, and used orbital elements to describe the orientation of the third body's orbit. In contrast, we fix the orientation of the third body's orbit, and use orbital vectors/elements to describe the binary's orientation. In practice, this implies $i_{\mathrm{HR96}} = i$, $\omega_{\mathrm{HR96}} = -\Omega$, and $\Omega_{\mathrm{HR96}}=-\omega$, where elements with the subscript HR96 refer to \citet{1996MNRAS.282.1064H}, and elements without subscripts refer to the elements adopted here.

\subsection{Equations of motion}
\label{sect:prev:eom}
We expand the Hamiltonian of the 3-body system in terms of the ratio $x\equiv r/R$ of the separation of the binary, $r$, to the separation of the third body relative to the binary center of mass, $R$. To lowest order in $x$, i.e., $x^2$ (quadrupole order), and after averaging over the binary mean motion assuming a Kepler orbit, the equations of motion for the binary orbital elements read (e.g., \citealt{2018MNRAS.476.4139H})
\begin{subequations}
\label{eq:EOM_SA_gen}
\begin{align}
\frac{\mathrm{d}\ve{e}}{\mathrm{d} \theta} &= \epssa (1+\eper \cos \theta) \left [-3 \left(\ve{\j}\times \ve{e}\right) - \frac{3}{2} \left (\ve{\j} \cdot \unit{R} \right) \left(\ve{e} \times \unit{R} 
\right )+ \frac{15}{2} \left( \ve{e}\cdot \unit{R} \right ) \left ( \ve{\j} \times \unit{R} \right )\right ]; \\
\frac{\mathrm{d}\ve{\j}}{\mathrm{d} \theta} &= \epssa (1+\eper \cos \theta) \left [-\frac{3}{2}  \left ( \ve{\j} \cdot \unit{R} \right ) \left( \ve{\j} \times \unit{R} \right ) + \frac{15}{2} \left ( \ve{e} \cdot \unit{R} \right ) \left ( \ve{e} \times \unit{R} \right ) \right ].
\end{align}
\end{subequations}
Here, $\theta$ is the true anomaly of the perturber's (parabolic or hyperbolic) orbit, and the small parameter
\begin{align}
\label{eq:epssa}
\epssa \equiv \left [ \frac{\mper^2}{m(m+\mper)} \left ( \frac{a}{\qper} \right )^3 \left(1+\eper \right )^{-3} \right ]^{1/2}.
\end{align}
Generally, $-\Leper<\theta<\Leper$, where
\begin{align}
\label{eq:theta0_def}
\Leper\equiv \arccos\left ( -\frac{1}{\eper} \right ).
\end{align}

Assuming that $\unit{R}$ lies in the $(x,y)$-plane, we can write $\unit{R}=\cos\theta\, \unit{x} + \sin\theta\,\unit{y}$ whereas for the binary, $\ve{e} = e_x\,\unit{x} + e_y\,\unit{y} +e_z\,\unit{z}$ and $\ve{\j} = \j_x\,\unit{x} + \j_y\,\unit{y} +\j_z\,\unit{z}$. Therefore, equations~(\ref{eq:EOM_SA_gen}) can be written in the explicit form
\begin{subequations}
\label{eq:EOM_SA_spec}
\begin{align}
\nonumber \frac{\mathrm{d}\ve{e}}{\mathrm{d} \theta} &= -\frac{3}{2} \epssa (1+\eper \cos \theta)  \Biggl [ \frac{1}{2} \left \{ 3e_z \j_y + e_y \j_z + \left(e_z \j_y - 5e_y \j_z \right) \cos2\theta + (-e_z \j_x + 5e_x \j_z) \sin2 \theta \right \}  \unit{x} \\
\nonumber &\quad + \left \{ -2e_z \j_x+2e_x\j_z + \left(e_z\j_x-5e_x \j_z\right) \cos^2\theta + (e_z\j_y-5e_y \j_z) \cos\theta \sin\theta \right \}  \unit{y}  \\
&\quad + \left \{ -e_y \j_x + e_x \j_y + 2(e_y \j_x + e_x \j_y) \cos 2\theta + \left(-2e_x \j_x + 2e_y \j_y \right ) \sin 2\theta \right \} \unit{z} \Biggl ]; \\
\nonumber \frac{\mathrm{d}\ve{\j}}{\mathrm{d} \theta} &= \frac{3}{2} \epssa (1+\eper \cos \theta)  \Biggl [ \left \{ \left(-5 e_x e_z + \j_x \j_z\right) \cos\theta + (-5 e_y e_z + \j_y \j_z) \sin\theta \right \} \sin \theta \, \unit{x}  \\
\nonumber &\quad + \left \{ \left(5 e_x e_z - \j_x \j_z\right) \cos\theta + (5 e_y e_z - \j_y \j_z) \sin\theta \right \} \cos \theta \, \unit{y} \\
&\quad + \frac{1}{2} \left \{ (-10 e_x e_y+2\j_x \j_z) \cos 2\theta + \left(5e_x^2-5e_y^2-\j_x^2+\j_y^2 \right ) \sin 2\theta \right \} \unit{z} \Biggl ].
\end{align}
\end{subequations}

We will refer to equations~(\ref{eq:EOM_SA_gen}) and (\ref{eq:EOM_SA_spec}) as the SA equations of motion: averaged over the binary, but still explicitly a function of the perturber's orbital phase $\theta$.

\subsection{First-order equations}
\label{sect:prev:DA}
The SA equations of motion constitute a system of ODEs and can be solved numerically. In addition, as in previous works (\citealt{1996MNRAS.282.1064H,2018MNRAS.476.4139H}), the SA equations can be integrated over time from minus to plus infinity, or equivalently, over $\theta$ with $-\Leper<\theta<\Leper$, to obtain the changes in $\ve{e}$ and $\ve{\j}$. In the first-order (FO) approximation, the orbital vectors are assumed to be constant during the integration, i.e.,\begin{subequations}
\label{eq:FO_def}
\begin{align}
\label{eq:FO_def:e}
\nonumber(\Delta \ve{e})_\FO &\equiv \int_{-\Leper}^{\Leper} \mathrm{d} \theta \, \frac{\mathrm{d}\ve{e}}{\mathrm{d} \theta}(\ve{e}_0,\ve{\j}_0) = \epssa \frac{1}{2\eper} \Biggl [ \sqrt{1-1/\eper^2} \left \{ e_z \j_y \left(1-10\eper^2\right) + e_y \j_z \left(-5+2\eper^2 \right) \right \} - 3 \eper \left(3e_z \j_y+e_y\j_z \right ) \Leper \Biggl ] \unit{x} \\
\nonumber &\quad + \epssa \frac{1}{2\eper} \Biggl [ \sqrt{1-1/\eper^2} \left \{ e_z \j_x \left(1+8\eper^2\right) + e_x \j_z \left(-5+8\eper^2 \right) \right \} + 3 \eper \left(3e_z \j_x+e_x\j_z \right ) \Leper \Biggl ] \unit{y} \\ 
&\quad + \epssa \frac{1}{\eper} \Biggl [ \sqrt{1-1/\eper^2} \left \{ e_y \j_x \left (2+\eper^2 \right ) + e_x \j_y \left(2-5\eper^2 \right ) \right \} + 3 \eper L (e_y \j_x - e_x \j_y) \Biggl ] \unit{z}; \\
\label{eq:FO_def:j}
\nonumber (\Delta \ve{\j})_\FO &\equiv \int_{-\Leper}^{\Leper} \mathrm{d} \theta \, \frac{\mathrm{d}\ve{\j}}{\mathrm{d} \theta}(\ve{e}_0,\ve{\j}_0) = -\epssa \frac{1}{2\eper} (5e_y e_z - \j_y \j_z) \Biggl [\sqrt{1-1/\eper^2} \left(1+2\eper^2\right ) + 3\eper \Leper \Biggl ] \unit{x} \\
\nonumber &\quad + \epssa \frac{1}{2\eper} (5e_x e_z - \j_x \j_z) \Biggl [\sqrt{1-1/\eper^2} \left(-1+4\eper^2\right ) + 3\eper \Leper \Biggl ] \unit{y} \\
&\quad + \epssa \frac{1}{\eper} (5e_x e_y - \j_x\j_y) \sqrt{1-1/\eper^2} \left(1-\eper^2\right) \unit{z}.
\end{align}
\end{subequations}
In equations~(\ref{eq:FO_def}), the integrands are evaluated at the {\it initial} values of the eccentricity and angular-momentum vectors, and the latter are assumed to be constant throughout the perturber's passage (the components of $\ve{e}$ and $\ve{\j}$ in the explicit expressions in equations~\ref{eq:FO_def} should be interpreted as the components of the initial vectors; for brevity, we omitted the subscript `0'). In other words, in the FO approximation, the binary is not allowed to respond to the instantaneous perturbation of the third body.

In the limit when the change in the eccentricity vector is small, equation~(\ref{eq:FO_def:e}) yields the following more manageable expression for the scalar eccentricity change,
\begin{align}
\label{eq:Delta_e_FO_approx}
(\Delta e)_\FO \simeq \unit{e}\cdot (\Delta \ve{e})_\FO = \frac{5}{2} \epssa \frac{1}{e \eper} \Biggl [ \sqrt{1-1/\eper^2} \left \{ \left(1+2\eper^2\right) e_y e_z \j_x + \left(1-4\eper^2\right) e_x e_z \j_y - 2 \left(1-\eper^2 \right ) e_x e_y \j_z \right \} + 3 \eper \Leper e_z (e_y \j_x - e_x \j_y) \Biggl ].
\end{align}
It can be shown that equation~(\ref{eq:Delta_e_FO_approx}) is equivalent to the quadrupole-order result of \citeauthor{1996MNRAS.282.1064H} (\citeyear{1996MNRAS.282.1064H}; see also Appendix A2.2 of \citealt{2018MNRAS.476.4139H}).

\section{Analytic second-order computation}
\label{sect:cor}
The FO approximation works well in the limit when the changes in the binary's state are small. However, in some cases, the change in $\ve{e}$ and $\ve{\j}$ can be significant compared to the initial values (for example, when the initial eccentricity is already very high). In this case, it is no longer justified to assume that $\ve{e}$ and $\ve{\j}$ are constant as in equations~(\ref{eq:FO_def}), i.e., FO perturbation theory is insufficient (see also \citealt{2018MNRAS.476.4139H}). 

To derive improved expressions, we adopt a similar approach as \citet{2016MNRAS.458.3060L}. In short, we develop Fourier expansions of the SA equations of motion. The orbital vectors, $(\ve{e},\ve{\j})$, are also written in the form of Fourier expansions. By plugging the Fourier expansions of $\ve{e}$ and $\ve{\j}$ into the equations of motion, we obtain explicit relations for the Fourier coefficients associated with $\ve{e}$ and $\ve{\j}$. Subsequently, we integrate the equations of motion, now taking into account the leading order Fourier coefficients. The result consists of the original FO term, plus new additional terms to second order (SO). In other words, using Fourier expansions, we compute the instantaneous response of the binary to the perturber, and use the updated expressions to more accurately calculate the net effects on the binary elements.

\subsection{Fourier expansions of the SA equations of motion}
\label{sect:cor:four}
\subsubsection{Calculation}
\label{sect:cor:four:calc}

First, we develop the Fourier series expansions of equations~(\ref{eq:EOM_SA_spec}), i.e.,
\begin{subequations}
\label{eq:eom_four}
\begin{align}
\frac{\mathrm{d}\ve{e}}{\mathrm{d} \theta} &= \epssa \Biggl [ \etz + \sum_{l=1}^{\lmax} \left \{ \etcl \cos \left ( \frac{l\pi}{\Leper} \theta \right) + \etsl \sin \left ( \frac{l\pi}{\Leper} \theta \right ) \right \}\Biggl ]; \\
\frac{\mathrm{d}\ve{\j}}{\mathrm{d} \theta} &= \epssa \Biggl [ \jtz + \sum_{l=1}^{\lmax} \left \{ \jtcl \cos \left ( \frac{l\pi}{\Leper} \theta \right) + \jtsl \sin \left ( \frac{l\pi}{\Leper} \theta \right ) \right \}\Biggl ],
\end{align}
\end{subequations}
with the Fourier coefficients given by
\begin{align}
\label{eq:gen_fourier}
\etz &= \frac{1}{2} \frac{1}{\epssa} \frac{1}{\Leper} \int_{-\Leper}^{\Leper} \mathrm{d} \theta\, \frac{\mathrm{d}\ve{e}}{\mathrm{d} \theta} (\ve{e}_0,\ve{\j}_0); \quad \etcl = \frac{1}{\epssa} \frac{1}{\Leper} \int_{-\Leper}^{\Leper} \mathrm{d} \theta\, \cos \left ( \frac{l\pi}{L} \theta \right ) \frac{\mathrm{d}\ve{e}}{\mathrm{d} \theta} (\ve{e}_0,\ve{\j}_0); \quad \etsl = \frac{1}{\epssa} \frac{1}{\Leper} \int_{-\Leper}^{\Leper} \mathrm{d} \theta\, \sin \left ( \frac{l\pi}{\Leper} \theta \right ) \frac{\mathrm{d}\ve{e}}{\mathrm{d} \theta} (\ve{e}_0,\ve{\j}_0); \\
\jtz &= \frac{1}{2} \frac{1}{\epssa} \frac{1}{\Leper} \int_{-\Leper}^{\Leper} \mathrm{d} \theta\, \frac{\mathrm{d}\ve{\j}}{\mathrm{d} \theta} (\ve{e}_0,\ve{\j}_0); \quad \jtcl = \frac{1}{\epssa} \frac{1}{\Leper} \int_{-\Leper}^{\Leper} \mathrm{d} \theta\, \cos \left ( \frac{l\pi}{\Leper} \theta \right ) \frac{\mathrm{d}\ve{\j}}{\mathrm{d} \theta} (\ve{e}_0,\ve{\j}_0); \quad \jtsl = \frac{1}{\epssa} \frac{1}{\Leper} \int_{-\Leper}^{\Leper} \mathrm{d} \theta\, \sin \left ( \frac{l\pi}{\Leper} \theta \right ) \frac{\mathrm{d}\ve{\j}}{\mathrm{d} \theta} (\ve{e}_0,\ve{\j}_0).
\end{align}
Here, the SA equations of motion in the integrands are to be evaluated at constant $\ve{e}$ and $\ve{\j}$ (i.e., the initial values). Explicit expressions for these Fourier coefficients are given in Appendix~\ref{app:fourier}. Note that, trivially, 
\begin{subequations}
\begin{align}
\etz &= \frac{(\Delta \ve{e})_\FO}{2\epssa \Leper}; \\
\jtz &= \frac{(\Delta \ve{\j})_\FO}{2\epssa \Leper}.
\end{align}
\end{subequations}

The summations in equations~(\ref{eq:eom_four}) run from $l=1$ to $l=\lmax$. In the case of a bound third body \citep{2016MNRAS.458.3060L}, the Fourier coefficients vanish identically for $l\geq4$. It turns out that, in our case of an unbound third body, the Fourier coefficients do {\it not} vanish for $l \geq 4$. In principle, an infinite number of terms should be included. However, in practice, it suffices to restrict to a finite $\lmax$. In our explicit expressions, we set $\lmax=3$ (see \S~\ref{sect:cor:four:test} below for the justification). 

Next, we assume that the binary orbital vectors can be written in terms of similar Fourier series but with the addition of a linear term, i.e.,
\begin{subequations}
\label{eq:ej_four}
\begin{align}
\ve{e} &= \ve{C}^{e}_1 + \ve{C}^{e}_2 \theta + \epssa \sum_{l=1}^{\lmax} \Biggl [ \ecl \cos \left ( \frac{l\pi}{\Leper} \theta \right) + \esl \sin \left ( \frac{l\pi}{\Leper} \theta \right ) \Biggl ] + \mathcal{O} \left ( \epssa^2 \right ); \\
\ve{\j} &= \ve{C}^{\j}_1 + \ve{C}^{\j}_2 \theta + \epssa \sum_{l=1}^{\lmax} \Biggl [ \jcl \cos \left ( \frac{l\pi}{\Leper} \theta \right) + \jsl \sin \left ( \frac{l\pi}{\Leper} \theta \right ) \Biggl ] + \mathcal{O} \left ( \epssa^2 \right ).
\end{align}
\end{subequations}
Here, $\ve{C}^{e}_1$, $\ve{C}^{e}_2$, $\ve{C}^{\j}_1$ and $\ve{C}^{\j}_2$ are constant vectors, and the Fourier coefficients $\ecl$, $\esl$, $\jcl$ and $\jsl$ associated with $\ve{e}$ and $\ve{\j}$ are understood to be functions of the initial values of $\ve{e}$ and $\ve{\j}$. We substitute equations~(\ref{eq:ej_four}) into equations~(\ref{eq:eom_four}) by differentiating the former with respect to $\theta$. Subsequently, we equate the coefficients of all terms (constant terms, and the sine and cosine terms) on both sides of the resulting equality. This procedure gives the following relations between the coefficients of the orbital vectors, $\ve{e}$ and $\ve{\j}$, and those of the equations of motion,
\begin{subequations}
\begin{align}
\ve{C}^{e}_2 &= \etz; \quad \ecl = - \frac{\Leper}{l\pi} \etcl+ \mathcal{O} \left ( \epssa \right ); \quad \esl = \frac{\Leper}{l\pi} \etsl + \mathcal{O} \left ( \epssa \right ); \\
\ve{C}^{\j}_2 &= \jtz; \quad \jcl = - \frac{\Leper}{l\pi} \jtcl + \mathcal{O} \left ( \epssa \right ); \quad \jsl = \frac{\Leper}{l\pi} \jtsl + \mathcal{O} \left ( \epssa \right ).
\end{align}
\end{subequations}
The necessity of the linear terms in equations~(\ref{eq:ej_four}) becomes clear by noting that without them, the $\theta$-derivatives of equations~(\ref{eq:ej_four}) would not contain a term independent of $\theta$, whereas such a term is present in equations~(\ref{eq:eom_four}). We remark that the linear term was omitted in \citet{2016MNRAS.458.3060L}.  

Equations~(\ref{eq:ej_four}) describe the instantaneous response of the binary to the perturber in terms of an infinite Fourier series. They can be written in the more compact form
\begin{subequations}
\label{eq:ej_four_comp}
\begin{align}
\ve{e} = \ve{e}_0 + \epssa \Biggl [ (\theta+\Leper) \, \etz - \ve{F}_e(-\Leper) + \ve{F}_e(\theta) \Biggl ]  + \mathcal{O} \left ( \epssa^2 \right ); \\
\ve{\j} = \ve{\j}_0 + \epssa \Biggl [ (\theta+\Leper) \, \jtz - \ve{F}_\j(-\Leper) + \ve{F}_\j(\theta) \Biggl ] + \mathcal{O} \left ( \epssa^2 \right ).
\end{align}
\end{subequations}
Here, $\ve{e}_0$ and $\ve{\j}_0$ are the initial state vectors, and $\ve{F}_e(\theta)$ and $\ve{F}_\j(\theta)$ are (known) functions of $\theta$, defined by
\begin{subequations}
\begin{align}
\ve{F}_e(\theta) \equiv \sum_{l=1}^{\lmax} \Biggl [ \ecl \cos \left ( \frac{l\pi}{\Leper} \theta \right) + \esl \sin \left ( \frac{l\pi}{\Leper} \theta \right ) \Biggl ]; \\
\ve{F}_\j(\theta) \equiv \sum_{l=1}^{\lmax} \Biggl [ \jcl \cos \left ( \frac{l\pi}{\Leper} \theta \right) + \jsl \sin \left ( \frac{l\pi}{\Leper} \theta \right ) \Biggl ].
\end{align}
\end{subequations}

\subsubsection{Testing the Fourier expansions}
\label{sect:cor:four:test}

\begin{figure}
\center
\includegraphics[scale = 0.44, trim = 10mm 10mm 8mm 0mm]{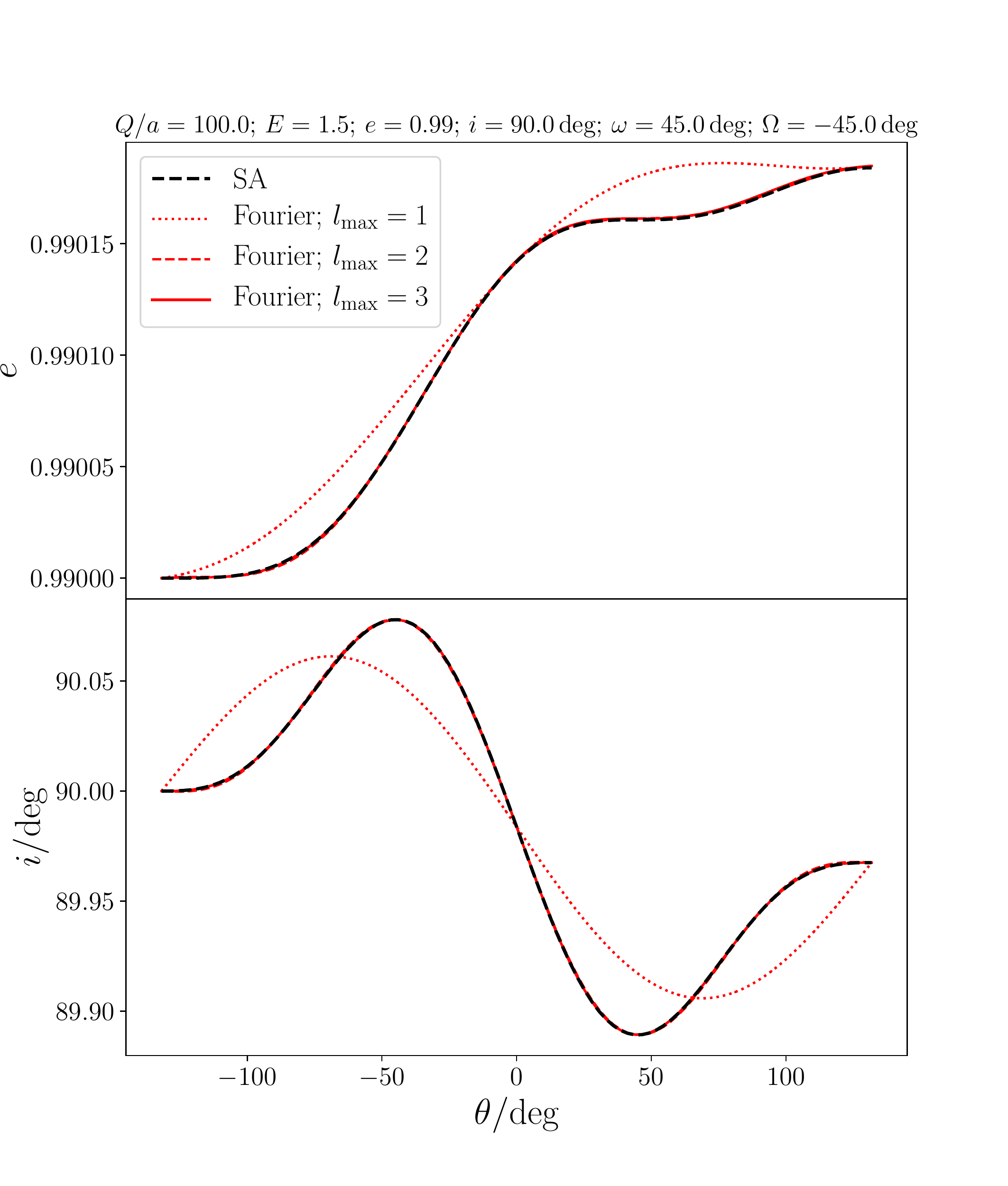}
\includegraphics[scale = 0.44, trim = 10mm 10mm 8mm 0mm]{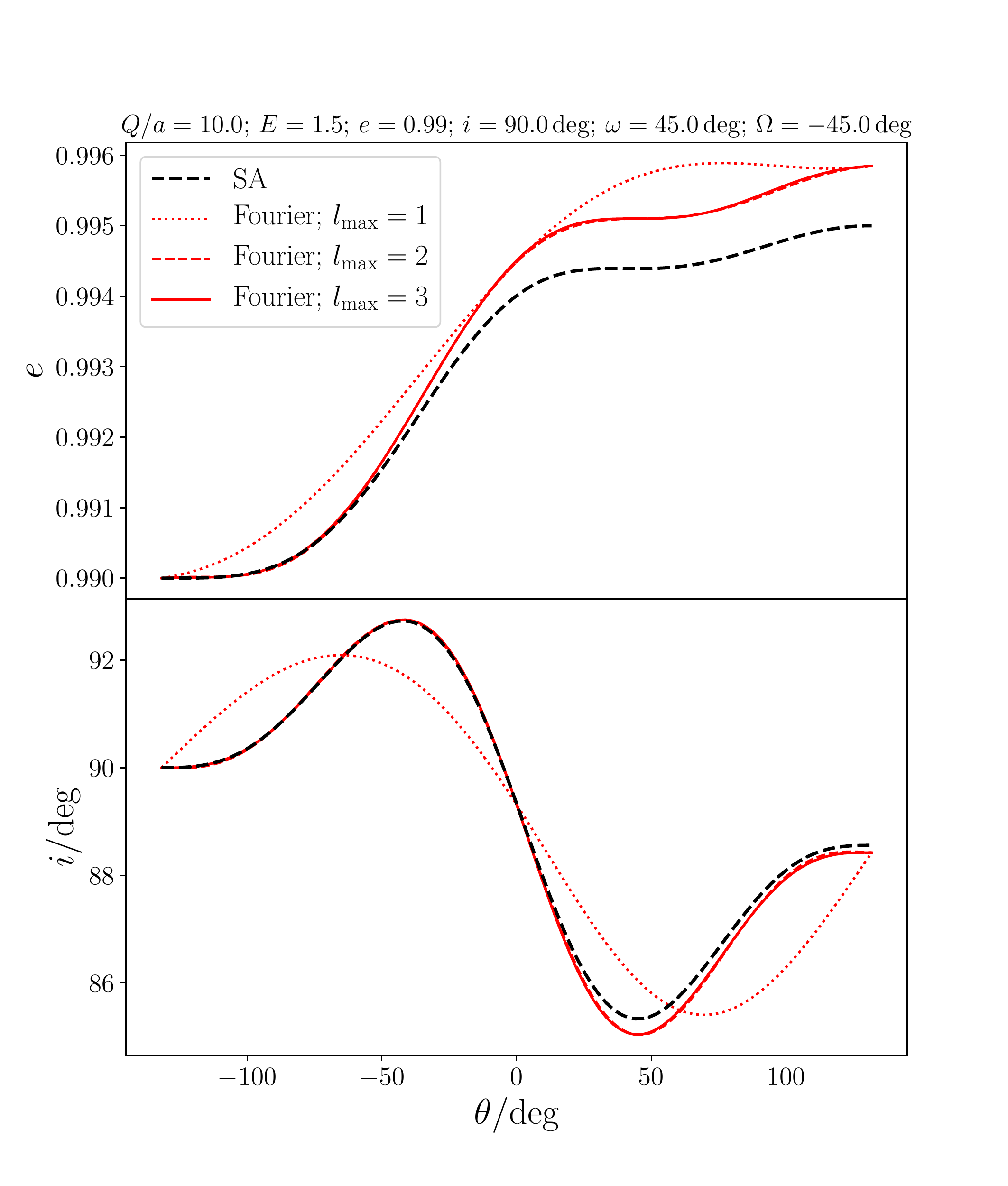}
\caption { Eccentricity (top panels) and inclination (bottom panels) of a binary perturbed by a third body as a function of the third body's orbital phase $\theta$. Fixed parameters are $m_1=m_2=\mper$, $\eper=1.5$, $e=0.99$, $i=90^\circ$, $\omega=45^\circ$, and $\Omega=-45^\circ$. In the left (right)-hand panels, $Q/a=100$ ($Q/a=10$), corresponding to $\epssa \simeq 1.0\times 10^{-4}$ ($\epssa \simeq 3.3\times 10^{-3}$). Black solid lines show $e$ and $i$ by numerically integrating the SA equations of motion, i.e., equations~(\ref{eq:EOM_SA_spec}); red lines correspond to the Fourier series in equations~(\ref{eq:ej_four_comp}), where we take $\lmax$ to be either 1, 2 or 3 and shown with the dotted, dashed and solid lines, respectively. }
\label{fig:fourier}
\end{figure}

To demonstrate the correctness and usefulness of the Fourier expansions, we show in \F~\ref{fig:fourier} the eccentricity and inclination of a binary perturbed by a third body as a function of the third body's orbital phase $\theta$. We set $m_1=m_2=\mper$, $\eper=1.5$, $e=0.99$, $i=90^\circ$, $\omega=45^\circ$, and $\Omega=-45^\circ$. In the left (right)-hand panels, $Q/a=100$ ($Q/a=10$), corresponding to $\epssa \simeq 1.0\times 10^{-4}$ ($\epssa \simeq 3.3\times 10^{-3}$). The black solid lines show $e$ and $i$ obtained by numerically integrating the SA equations of motion, i.e., equations~(\ref{eq:EOM_SA_spec}). Within the limit of the SA and quadrupole-order approximation, these solutions can be considered to be exact. Red lines correspond to the Fourier series in equations~(\ref{eq:ej_four_comp}), where we take $\lmax$ to be either 1, 2 or 3. 

In the case of $Q/a=100$, the Fourier series accurately describe $\ve{e}(\theta)$ and $\ve{\j}(\theta)$, provided that $\lmax>1$. There are no noticeable differences between $\lmax=2$ and $\lmax=3$. For $Q/a=10$ (i.e., larger $\epssa$), the Fourier series still give a reasonable approximation of the true $\ve{e}(\theta)$ and $\ve{\j}(\theta)$, although there are significant differences, in particular for $\theta \gtrsim 0^\circ$. The discrepancy in the $Q/a=10$ case between the numerical solutions and the Fourier series can be traced to the fact that we included terms of order $\epssa$ in equations~(\ref{eq:ej_four_comp}), and ignored terms of order $\epssa^2$. Evidently, higher-order terms become increasingly important as $\epssa$ increases. Note that convergence with respect to $\epssa$ is distinct from convergence with respect to $l$. In the case of $Q/a=10$, the series in $l$ still converges for $\lmax>1$.

\subsection{Calculating corrections to second order in $\epssa$}
\label{sect:cor:calc}
\subsubsection{General case: hyperbolic orbits}
\label{sect:cor:calc:gen}

The next step is to substitute $\ve{e}$ and $\ve{\j}$ up to and including first order in $\epssa$ into the equations of motion, equations~(\ref{eq:eom_four}), where the Fourier coefficients $\etz$, $\etcl$, $\etsl$, $\jtz$, $\jtcl$, and $\jtsl$ are now all evaluated using equations~(\ref{eq:ej_four_comp}), and subsequently integrating over $\theta$ from $-\Leper$ to $\Leper$. 

To first order in $\epssa$, this procedure simply yields 
\begin{subequations}
\begin{align}
(\Delta \ve{e})_{\epssa} &= \epssa 2\Leper \, \etz = (\Delta \ve{e})_\FO \equiv \epssa \fe; \\
\quad (\Delta \ve{\j})_{\epssa} &= \epssa 2\Leper \, \jtz = (\Delta \ve{\j})_\FO \equiv \epssa \fj,
\end{align}
\end{subequations}
i.e., we recover the FO result (note that, to zeroth order in $\epssa$, $\int_{-\Leper}^{\Leper} \mathrm{d} \theta \, \etcl \cos [l\pi \theta/\Leper] = 0$, and similarly for the sine terms and the other Fourier coefficients). For future convenience, we defined the two functions
\begin{align}
\fe(\ve{e}_0,\ve{\j}_0,\eper) \equiv 2 \Leper \etz(\ve{e}_0,\ve{\j}_0,\eper); \quad \fj(\ve{e}_0,\ve{\j}_0,\eper) \equiv 2 \Leper \jtz(\ve{e}_0,\ve{\j}_0,\eper).
\end{align}

To second order in $\epssa$, the general result is (much) more complicated. Specifically,
\begin{subequations}
\begin{align}
(\Delta \ve{e})_{\epssa^2} &= \epssa^2 \Biggl [\gei(\ve{e}_0,\ve{\j}_0,\eper) + \geii (\ve{e}_0,\ve{\j}_0,\eper) \Biggl ] \equiv \epssa^2 \ge(\ve{e}_0,\ve{\j}_0,\eper); \\
(\Delta \ve{\j})_{\epssa^2} &= \epssa^2 \Biggl [\gji (\ve{e}_0,\ve{\j}_0,\eper) +\gjii (\ve{e}_0,\ve{\j}_0,\eper) \Biggl ] \equiv \epssa^2 \gj(\ve{e}_0,\ve{\j}_0,\eper).
\end{align}
\end{subequations}
Here, $\ge$ and $\gj$, which are composed of $\gei+\geii$ and $\gji+\gjii$ respectively, are known functions of $\ve{e}_0$, $\ve{\j}_0$, and $\eper$. The functions $\gei$ and $\gji$ are easily obtained through
\begin{align}
\label{eq:gei_gji}
\gei (\ve{e}_0,\ve{\j}_0,\eper)= \fe(\ve{e}',\ve{\j}',\eper); \quad \gji(\ve{e}_0,\ve{\j}_0,\eper) = \fj(\ve{e}',\ve{\j}',\eper).
\end{align}
Here, $\ve{e}'$ and $\ve{\j}'$ are the $\theta$-independent terms of equations~(\ref{eq:ej_four_comp}), i.e.,
\begin{align}
\label{eq:ej_sub}
\ve{e}' = \ve{e}_0 - \epssa \left [\etz \Leper - \ve{F}_e(\Leper) \right ]; \quad \ve{\j}' = \ve{\j}_0 - \epssa \left [\jtz \Leper - \ve{F}_\j(\Leper) \right ].
\end{align}
In other words, $\gei$ and $\gji$ can be obtained by taking the FO result (i.e., $\fe$ and $\fj$ for the eccentricity and angular-momentum vector, respectively), and replacing the initial state vectors with $\ve{e}'$ and $\ve{\j}'$ given by equations~(\ref{eq:ej_sub}). Note that this result is independent of the value of $\lmax$. The explicit substitution relations in equations~(\ref{eq:ej_sub}) are given in Appendix~\ref{app:SO}. The functions $\geii$ and $\gjii$ do depend on the assumed value of $\lmax$, and are generally very complicated. We therefore do not give the explicit expressions for $\geii$ and $\gjii$ here. Instead, we implemented all required functions in an easy-to-use script that is freely available and described below (see \S~\ref{sect:ex}).

In compact notation, the SO result can be written as
\begin{subequations}
\begin{align}
\Delta \ve{e} &= \epssa \fe + \epssa^2 \ge + \mathcal{O}\left(\epssa^3\right); \\
\Delta \ve{\j} &= \epssa \fj + \epssa^2 \gj + \mathcal{O}\left(\epssa^3\right).
\end{align}
\end{subequations}
In the limit of small changes, the scalar eccentricity change to second order in $\epssa$ is given by
\begin{align}
\label{eq:Delta_e_SO}
(\Delta e)_\SO \simeq \epssa (\unit{e} \cdot \fe) + \epssa^2 (\unit{e} \cdot \ge). 
\end{align}
Generally, the inclination change is given by
\begin{align}
\label{eq:Delta_i_gen}
\Delta i = \arccos \left ( \frac{\ve{\j}+\Delta \ve{\j}}{||\ve{\j}+\Delta \ve{\j}||}\cdot \unit{z} \right ) - \arccos \left ( \frac{\ve{\j}\cdot \unit{z}}{||\ve{\j}||} \right ).
\end{align}

\subsubsection{The parabolic limit} 
\label{sect:cor:calc:par}
In the parabolic limit ($\eper\rightarrow 1$), the SO expressions become much less complicated. Specifically,
\begin{subequations}
\label{eq:Delta_ej_vec_CDA_par}
\begin{align}
\nonumber \ge &= \frac{3}{16} \pi  \epssa^2 \left[75 e_x^2 e_y-6 \pi  \left(e_x \left(15 e_z^2-6 \j_y^2+\j_z^2\right)+6 e_y \j_x \j_y\right)+50 e_y^3+5 e_y \left(10 e_z^2+\j_x^2-10 \left(\j_y^2+2 \j_z^2\right)\right)+50 e_z \j_y \j_z\right] \unit{x} \\
\nonumber &\quad - \frac{3}{16} \pi  \epssa^2 \left[75 e_x^3+e_x \left(50 e_y^2+5 \j_x^2+36 \pi  \j_x \j_y-150 \j_z^2\right)+6 \pi  e_y \left(15 e_z^2-6 \j_x^2+\j_z^2\right)-10 \j_x (5 e_y \j_y+e_z \j_z)\right ] \unit{y} \\
&\quad +  \frac{3}{8} \pi  \epssa^2 \left[-6 \pi  e_z \left(5 e_x^2+5 e_y^2-3 \left(\j_x^2+\j_y^2\right)\right)-5 (5 e_x e_y e_z+15 e_x \j_y \j_z-9 e_y \j_x \j_z+5 e_z \j_x \j_y)+12 \pi  \j_z (e_x \j_x+e_y \j_y)\right ] \unit{z}; \\
\nonumber \gj &= \frac{3}{16} \pi  \epssa^2 \left[75 e_x^2 \j_y+60 \pi  e_x e_y \j_y-6 \pi  \j_x \left(10 e_y^2+15 e_z^2+\j_z^2\right)-50 e_y e_z \j_z+50 e_z^2 \j_y+5 \j_x^2 \j_y\right] \unit{x} \\
\nonumber &\quad -\frac{3}{16} \pi  \epssa^2 \left[15 e_x^2 (5 \j_x+4 \pi  \j_y)-10 e_x (6 \pi  e_y \j_x+5 e_y \j_y+15 e_z \j_z)+5 \j_x \left(10 e_y^2+\j_x^2-2 \j_z^2\right)+6 \pi  \j_y \left(15 e_z^2+\j_z^2\right)\right] \unit{y} \\
&\quad -\frac{15}{8} \pi  \epssa^2 \left[5 e_x e_y \j_z+5 e_x e_z \j_y+5 e_y e_z \j_x+\j_x \j_y \j_z\right] \unit{z}.
\end{align}
\end{subequations}
The above expressions imply that, in the limit of small changes and parabolic orbits, the scalar eccentricity change is given by
\begin{align}
\label{eq:Delta_e_CDA_par}
\nonumber &(\Delta e)_\SO(\ve{e},\ve{\j},\eper\rightarrow1) \simeq \epssa (\unit{e}\cdot \fe) + \epssa^2 (\unit{e}\cdot \ge) = \epssa \frac{15\pi}{2e} e_z (e_y \j_x - e_x \j_y) \\
\nonumber &\qquad + \epssa^2 \frac{3 \pi}{8e}  \biggl [25 \left(e_x e_y \left(\j_z^2-\j_y^2\right)-e_z \j_y (2 e_x \j_z+e_z \j_x)+e_y^2 \j_x \j_y+2 e_y e_z \j_x \j_z\right)-3 \pi  \biggl \{e_x^2 \left(25 e_z^2-6 \j_y^2+\j_z^2\right)+4 e_x \j_x (3 e_y \j_y-e_z \j_z)  \\
\nonumber &\qquad \qquad +e_y^2 \left(25 e_z^2-6 \j_x^2+\j_z^2\right)-4 e_y e_z \j_y \j_z-6 e_z^2 \left(\j_x^2+\j_y^2\right)\biggl \}\biggl ]\\
\nonumber &= \epssa \frac{15\pi}{4} e \sqrt{1-e^2} \sin 2\omega \sin^2i \\
\nonumber &\qquad + \epssa^2 \frac{3}{512} \pi e \biggl [ 100 \left(1-e^2\right) \sin 2 \omega  \biggl \{ \left (5 \cos i+3 \cos 3 i\right ) \cos 2 \Omega +6 \sin i \sin 2 i\biggl \}+4 \cos 2 i \biggl \{3 \pi  \left(81 e^2-56\right)+200 \left(1-e^2\right) \cos 2 \omega  \sin 2 \Omega \biggl \} \\
&\qquad \qquad +3 \pi  \biggl \{ 200 e^2 \sin ^4i \cos 4 \omega +8 \left(16 e^2+9\right) \sin ^2 2 i \cos 2 \omega +\left(39 e^2+36\right) \cos 4 i-299 e^2+124\biggl \}\biggl ].
\end{align}
In equation~(\ref{eq:Delta_e_CDA_par}), we also wrote the result in terms of the orbital elements. It can be shown that equation~(\ref{eq:Delta_e_CDA_par}) is consistent to order $\epssa$ with equation (8) of \citet{1996MNRAS.282.1064H}, using that $i_{\mathrm{HR96}} = i$ and $\Omega_{\mathrm{HR96}}=-\omega$, together with $\epssa$ evaluated at $\eper=1$ (cf. equation~\ref{eq:epssa}). The novel result of this paper is the SO term in equation~(\ref{eq:Delta_e_CDA_par}).

\section{Exemplifying the correctness and relevance of the second-order expressions}
\label{sect:ex}
In this section, we illustrate interesting behavior in the limit of high binary eccentricities, in which case the analytic FO expressions for the scalar eccentricity change can break down and a SO calculation is warranted. We demonstrate the correctness of our SO expressions by comparing to results obtained by numerically integrating the SA equations, and by directly integrating the 3-body equations of motion (i.e., without averaging over the binary's motion). 

We freely provide\footnote{\href{https://github.com/hamers/flybys}{https://github.com/hamers/flybys}} a simple \textsc{Python} script to compute the FO and SO expressions. In this script, which requires only the \textsc{numpy} and \textsc{scipy} libraries, we also include functionality to integrate the SA equations of motion numerically, and to carry out direct 3-body integration. The script can be used to quickly generate all the figures presented in this paper.

\subsection{Phase evolution}
\label{sect:ex:time}

\begin{figure}
\center
\includegraphics[scale = 0.46, trim = 8mm 10mm 8mm 0mm]{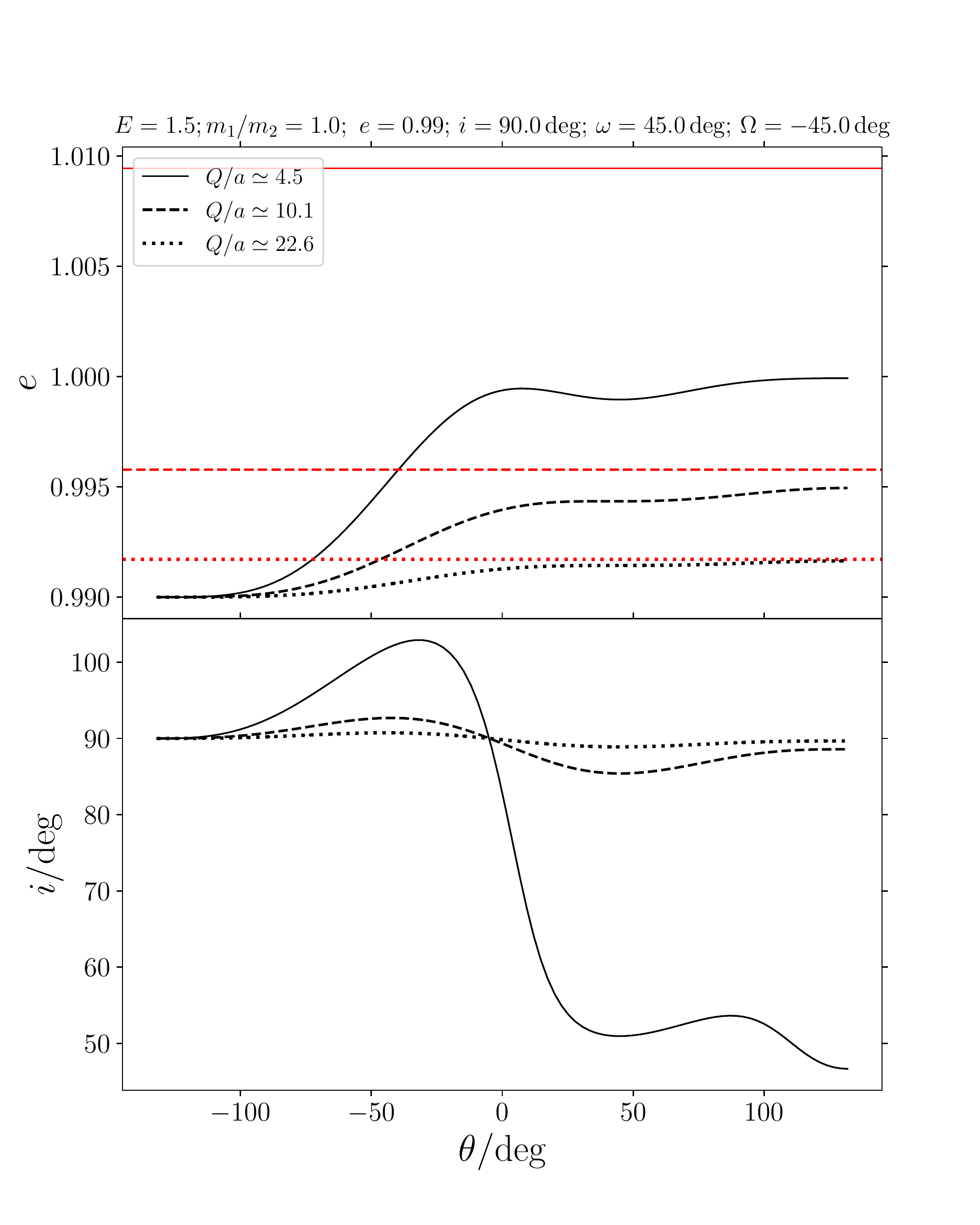}
\includegraphics[scale = 0.46, trim = 8mm 10mm 8mm 0mm]{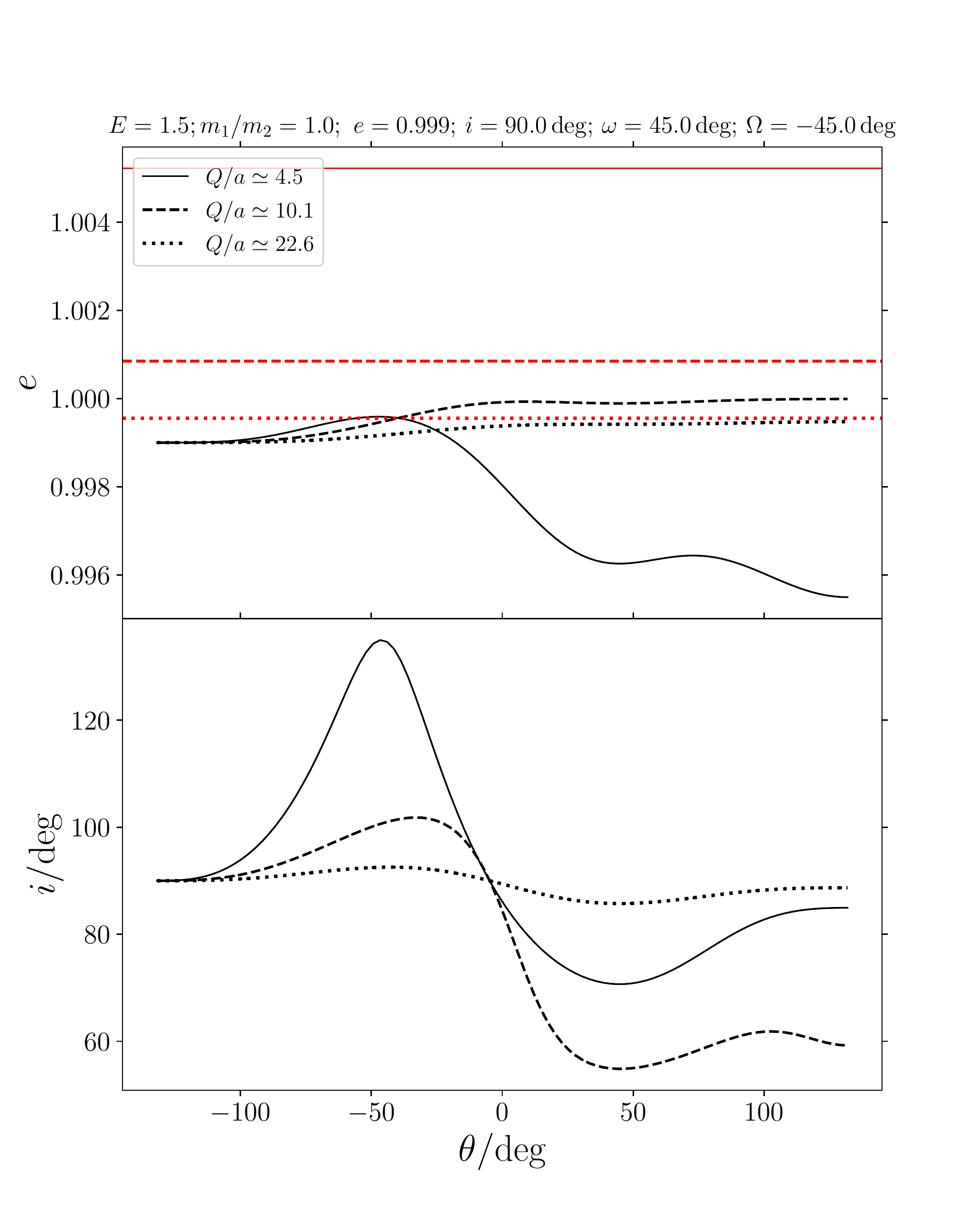}
\caption { Binary eccentricity (top panel) and inclination (bottom panel) as a function of the perturber's true anomaly $\theta$ obtained by numerically integrating the SA equations of motion, assuming $m_1=m_2=\mper$, $\eper=1.5$, $i=90^\circ$, $\omega=45^\circ$, and $\Omega=-45^\circ$. In the left (right)-hand panels, $e=0.99$ ($e=0.999)$. In each panel, three values of $Q/a$ are chosen indicated in the legends. The red horizontal lines show $e+\Delta e$ according to FO perturbation theory (equation~\ref{eq:Delta_e_FO_approx}), with the line styles corresponding to the legends. The FO clearly breaks down in some cases }
\label{fig:phase}
\end{figure}

In \F~\ref{fig:phase}, we show the binary eccentricity and inclination as a function of the perturber's true anomaly $\theta$ assuming $m_1=m_2=\mper$, $\eper=1.5$, $i=90^\circ$, $\omega=45^\circ$, and $\Omega=-45^\circ$. In the left (right)-hand panels, $e=0.99$ ($e=0.999)$. In each panel, three values of $Q/a$ are chosen, ranging from relatively large, ($Q/a\simeq 23$), to small ($Q/a \simeq 5$). These results were obtained by numerically integrating the SA equations of motion. The red horizontal lines show $e+\Delta e$ according to FO perturbation theory (equation~\ref{eq:Delta_e_FO_approx}). 

We first consider the case when the initial binary eccentricity is $e=0.99$ (left-hand panel in \F~\ref{fig:phase}). For relatively large $Q/a$, $Q/a\simeq 23$, the changes in eccentricity and inclination are small. The FO equation accurately describes the scalar eccentricity change. However, as $Q/a$ decreases, the agreement worsens, and for $Q/a\simeq 5$, the FO equation even yields $e+\Delta e>1$, which is clearly not physical (note that, according to both SA and direct integration, the binary remains bound with a high eccentricity). 

In the right-hand panel of \F~\ref{fig:phase}, the initial binary eccentricity is even higher, $e=0.999$ (all other parameters are the same as in the left-hand panel). As shown with the SA integrations, as $Q/a$ decreases, the maximum eccentricity reached during the perturber's passage increases. This continues until, at a critical value of $Q/a$, the maximum eccentricity reached during the passage approaches unity, and the binary subsequently `bounces back' to lower eccentricity. Consequently, the net scalar eccentricity change becomes {\it negative}. This phenomenon is also associated with a large change in the inclination, and is similar to the flipping of orbits associated with very high eccentricities in hierarchical systems such as triples (e.g., \citealt{2011ApJ...742...94L,2018MNRAS.481.4907G}), and higher-order systems (e.g., \citealt{2017MNRAS.470.1657H,2018MNRAS.474.3547G}). The FO equation fails to describe this `bounce' effect, and incorrectly predicts a net increase of the eccentricity, and which is $>1$.  

\subsection{Series of $Q/a$}
\label{sect:ex:series}

\begin{figure}
\center
\includegraphics[scale = 0.46, trim = 8mm 10mm 8mm 0mm]{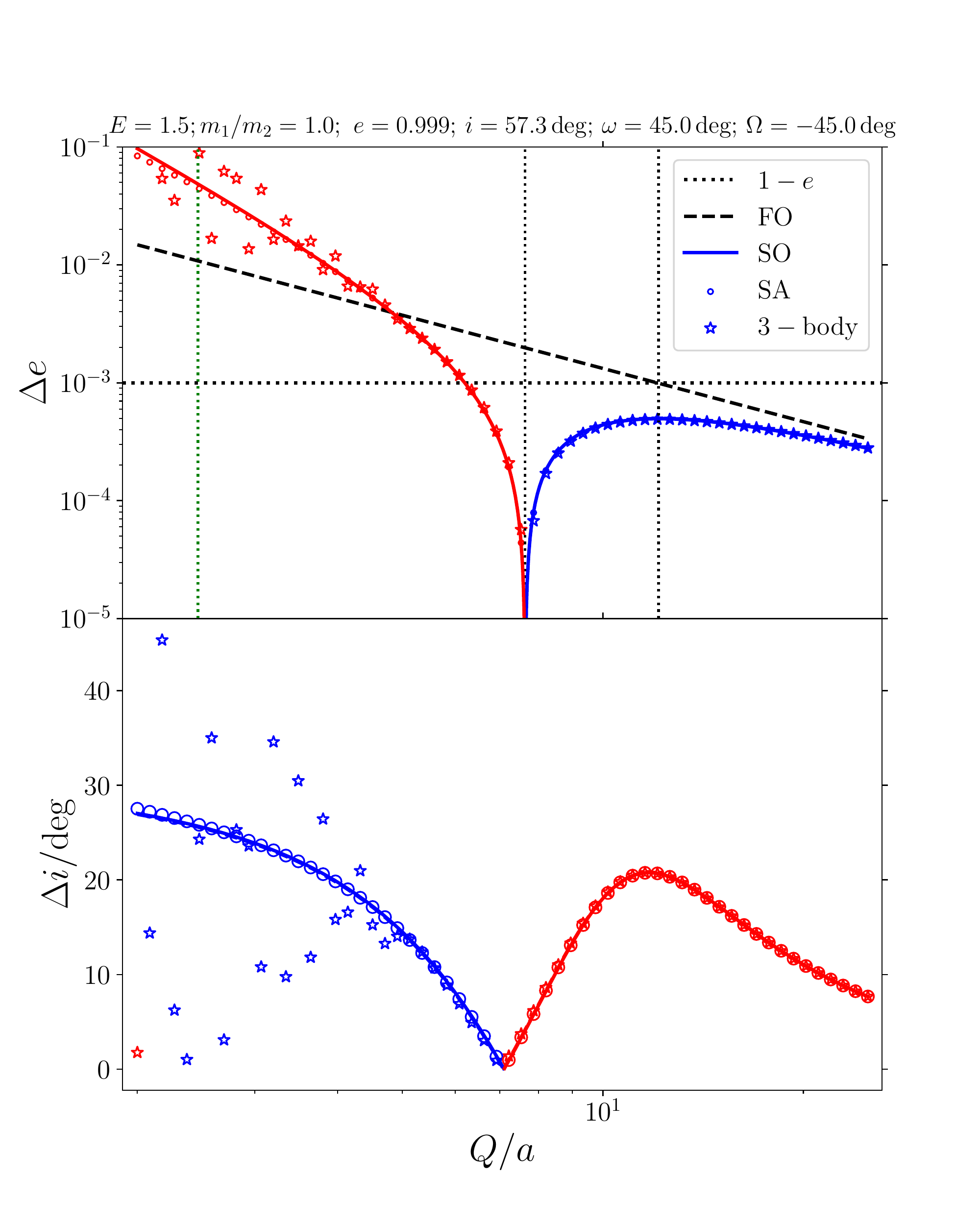}
\includegraphics[scale = 0.46, trim = 8mm 10mm 8mm 0mm]{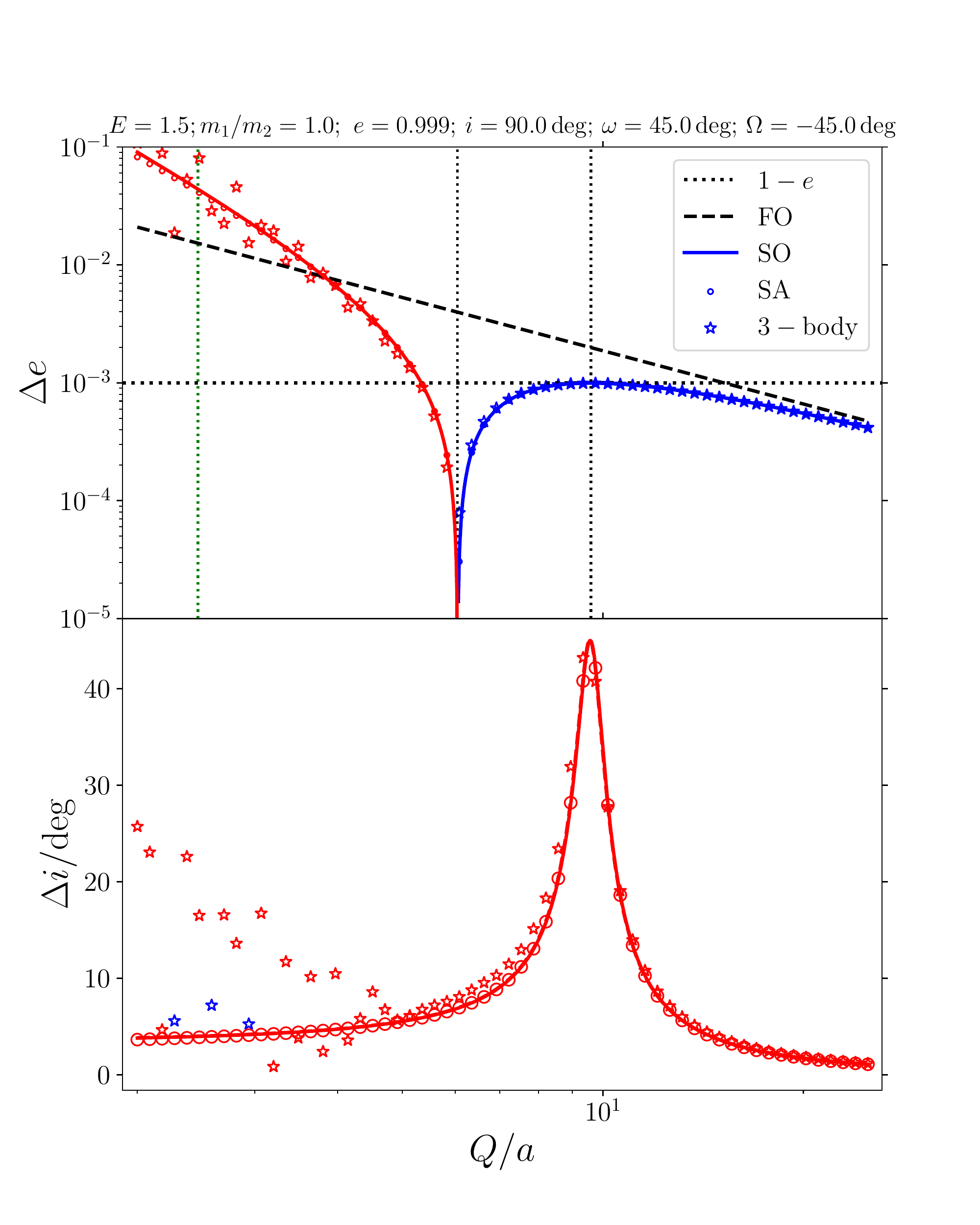}
\caption {Scalar binary eccentricity (top panels) and inclination (bottom panels) change as a function of $Q/a$. Fixed parameters are $m_1=m_2=\mper$, $e=0.999$, $\eper=1.5$, $\omega=45^\circ$, and $\Omega=-45^\circ$. In the left (right)-hand panels, the binary's inclination is $57.3^\circ$ ($90^\circ$). Results are shown from different techniques: analytic FO (black dashed lines), analytic SO (colored solid lines), numerical SA (colored open circles), and 3-body integration (colored stars). Blue (red) colors correspond to positive (negative) changes. The black horizontal dotted lines show $1-e$, the maximum allowed (positive) change. The green dotted vertical line shows $Q/a$ corresponding to $\mathcal{R}=0.5$ (see equation~\ref{eq:R_def}); encounters with $Q/a$ less than approximately this value are within the non-secular regime. The two vertical black dotted lines show the values of $Q/a$ for which we can analytically calculate the locations of the sign change and plateau in $\Delta e$ using the SO expressions (see \S~\ref{sect:ex:series:prop}). }
\label{fig:series1}
\end{figure}

\subsubsection{General behavior}
\label{sect:ex:series:gen}

As described above, for high initial binary eccentricities and assuming that for weak perturbations the binary's eccentricity is increased, for sufficiently strong perturbations (small $Q/a$), the binary eccentricity shows a `bounce' phenomenon, resulting in a net negative scalar eccentricity change. Here, we describe this effect more quantitatively and show that our SO results can be used to accurately describe the binary's eccentricity and angular-momentum changes in the associated parameter space.

In \F~\ref{fig:series1}, we show the scalar eccentricity change and inclination in the top and bottom panels, respectively, as a function of $Q/a$. The fixed parameters are $m_1=m_2=\mper$, $e=0.999$, $\eper=1.5$, $\omega=45^\circ$, and $\Omega=-45^\circ$. In the left (right)-hand panels, the binary's inclination is $57.3^\circ$ ($90^\circ$). We show analytic FO results (black dashed lines), analytic SO results (colored solid lines), numerical SA results (colored open circles), and 3-body integration results (colored stars). Blue (red) colors correspond to positive (negative) changes. 

The 3-body integrations were carried out by formulating the 3-body equations of motion into ODE form and solving this set of equations using the \textsc{odeint} routine from the \textsc{Python} package \textsc{scipy}. This routine is based on the \textsc{LSODA} Fortran integrator, which uses variable time steps to integrate the system of ODEs. \textsc{LSODA} automatically and dynamically switches between stiff and nonstiff methods; for stiff cases, it uses the backward differentiation formula method (with a dense or banded Jacobian), and for nonstiff cases it uses the Adams method. Error control within the solver is determined by the input parameters $\textsc{rtol}$ and $\textsc{atol}$, such that the error in each ODE variable $y_i$ is less than or equal to $\textsc{rtol} \times \mathrm{abs}(y_i)+\textsc{atol}$. We set the relative and absolute tolerance parameters to $\textsc{rtol} = 10^{-15}$ and $\textsc{atol} = 10^{-15}$, respectively. Typical relative energy errors were $10^{-5}$ (for small $Q/a$, corresponding to few binary orbits), to $10^{-3}$ (for large $Q/a$, corresponding to many binary orbits). 

We first describe the behavior shown in \F~\ref{fig:series1} according to the SA and 3-body integrations. For large $Q/a$, the SA and 3-body integrations agree well. As stated above, the 3-body energy errors are largest for large $Q/a$, but the good agreement with the SA results indicates that the 3-body integrations can still be trusted. As $Q/a$ decreases, $\Delta e$ starts to flatten, and reaches a local maximum, at $Q/a\sim 10$. This plateau in $\Delta e$ is associated with a large inclination change (see the bottom panels in \F\,\ref{fig:series1}). In the case when $i=90^\circ$, the plateau corresponds to $\Delta e = 1-e$, the maximum allowed positive change (shown with the black horizontal dotted lines). 

For smaller $Q/a$, $\Delta e$ decreases with increasing $Q/a$, and at a critical value of $Q/a$, $Q/a\sim 8$, $\Delta e$ changes sign. Subsequently, $|\Delta e|$ increases, with a different slope compared to before the sign change (in fact, this slope is $|\Delta e|\propto (Q/a)^{-3}$, as shown below in \S~\ref{sect:ex:series:prop}). At small $Q/a$, $Q/a\lesssim 5$, the SA and 3-body results start to differ significantly. This is the result of the breakdown of the SA approximation: for very small $Q/a$ ($Q/a$ approaching unity), the binary's motion is no longer much faster than the perturber's motion, therefore it is no longer justified to average over the binary motion. This is also supported by the green vertical dotted line, which shows the value of $Q/a$ for which $\mathcal{R}=0.5$ (see equation~\ref{eq:R_def}). The scatter in the 3-body results originates from the binary phase-dependence in this regime (we assumed a fixed initial binary phase of $\theta_{\mathrm{bin}}=1\,\mathrm{rad}$). 

The FO expression gives a good description at large $Q/a$, but clearly fails as $Q/a\lesssim 20$. The SO results, shown with the solid colored lines, agree well with the SA results for any $Q/a$ and, therefore, with the 3-body results, unless $Q/a \lesssim 5$. 

\begin{figure}
\center
\includegraphics[scale = 0.46, trim = 8mm 0mm 8mm 0mm]{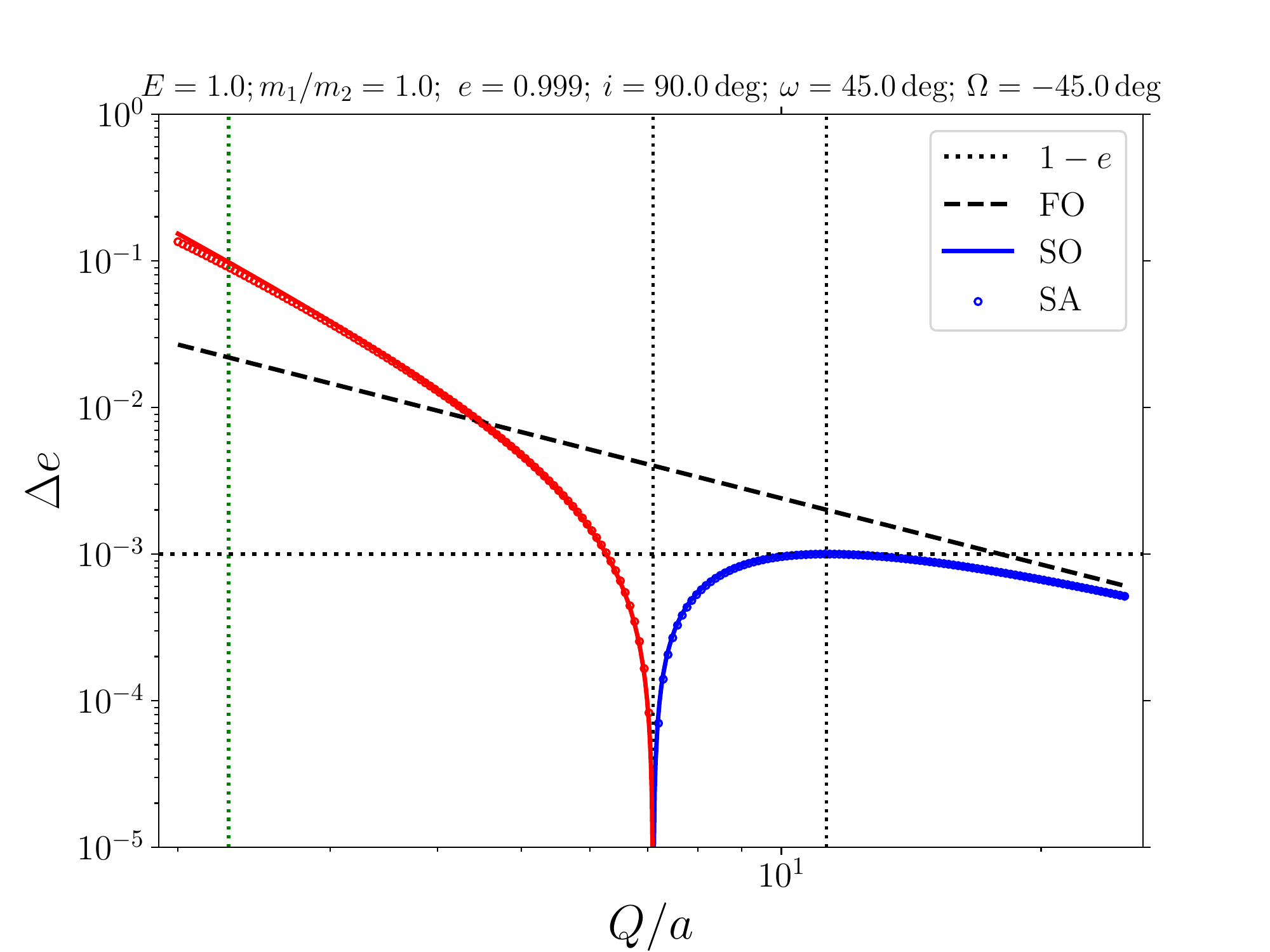}
\includegraphics[scale = 0.46, trim = 8mm 0mm 8mm 0mm]{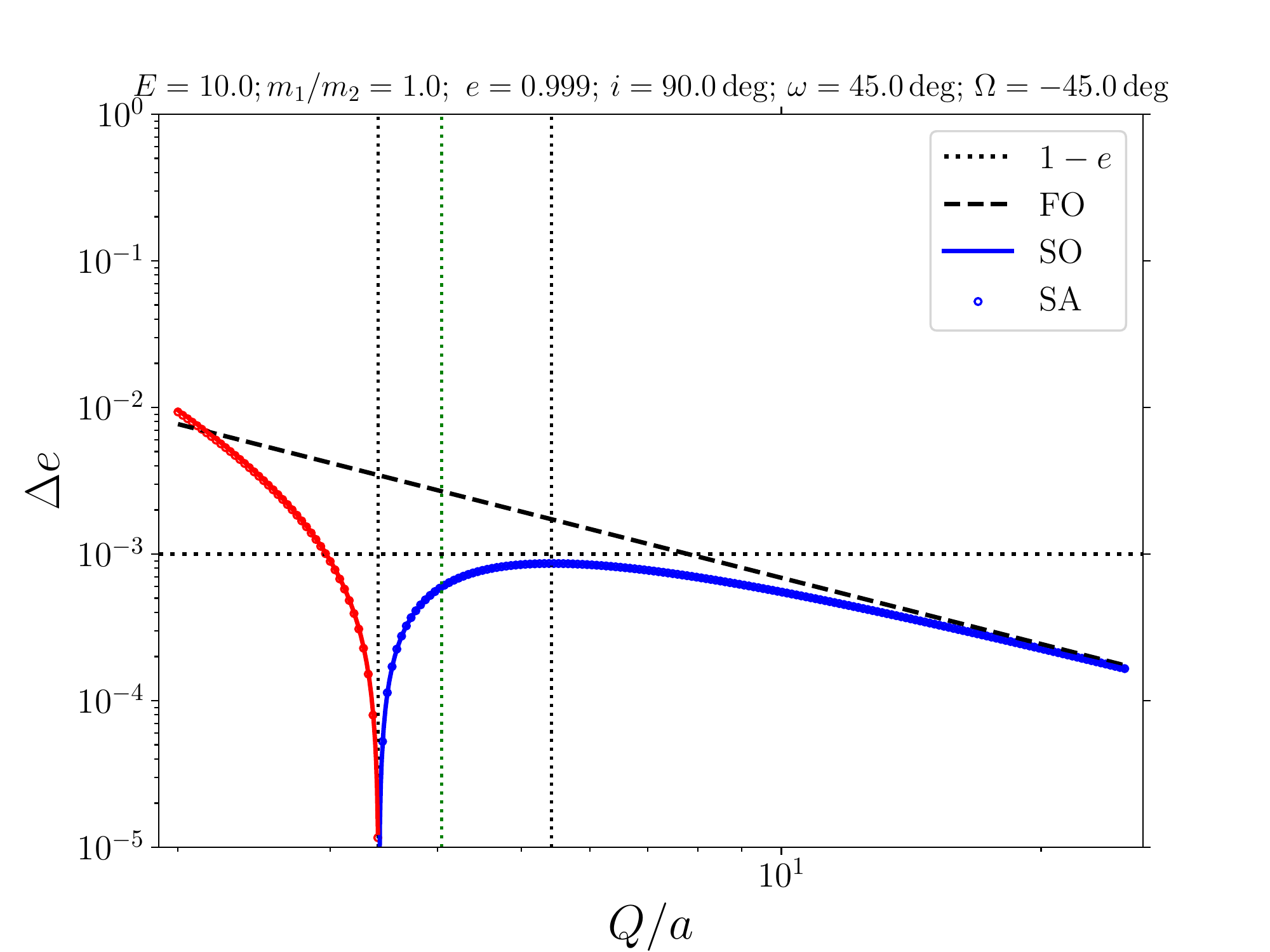}
\includegraphics[scale = 0.46, trim = 8mm 0mm 8mm 0mm]{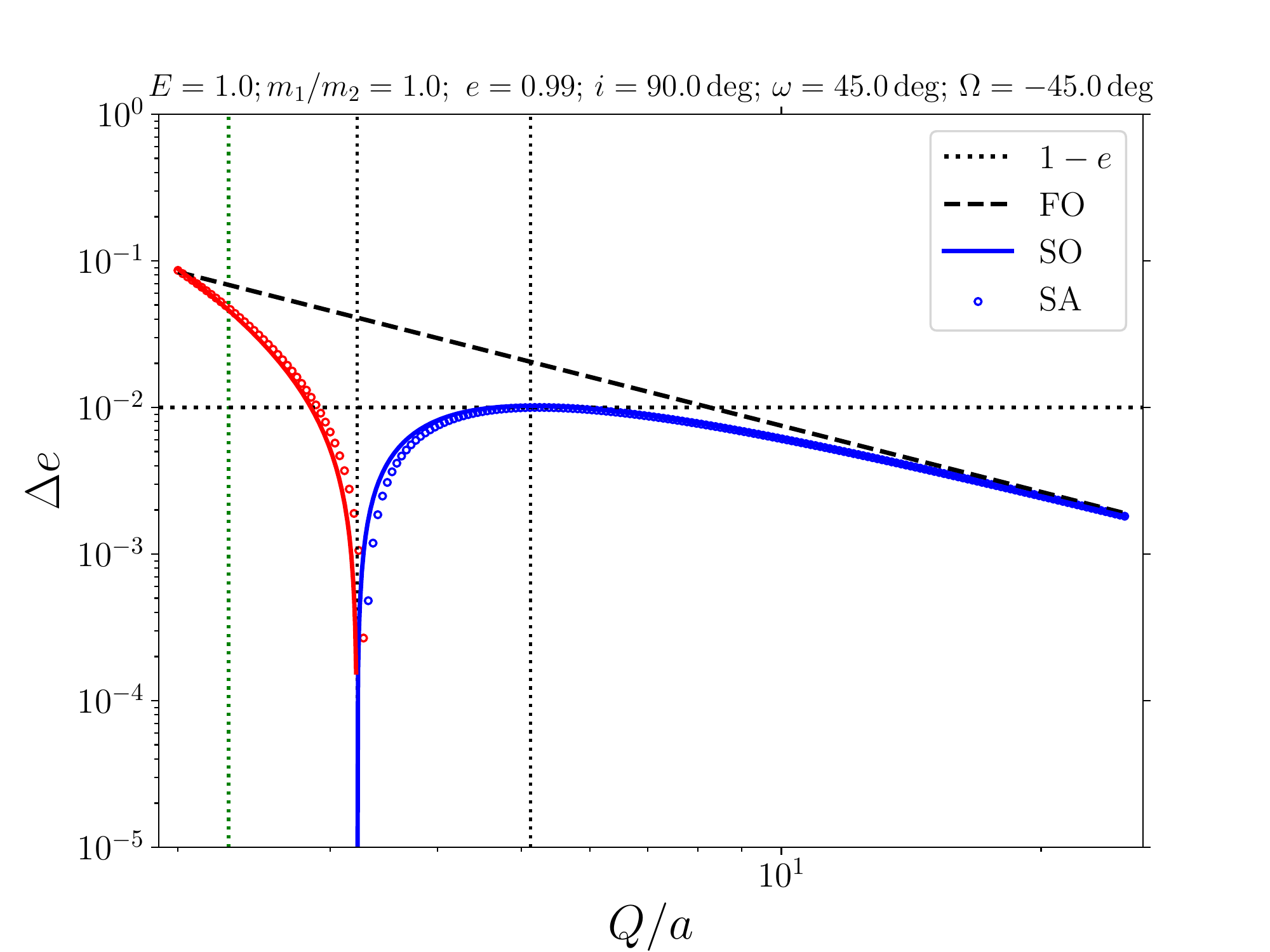}
\includegraphics[scale = 0.46, trim = 8mm 0mm 8mm 0mm]{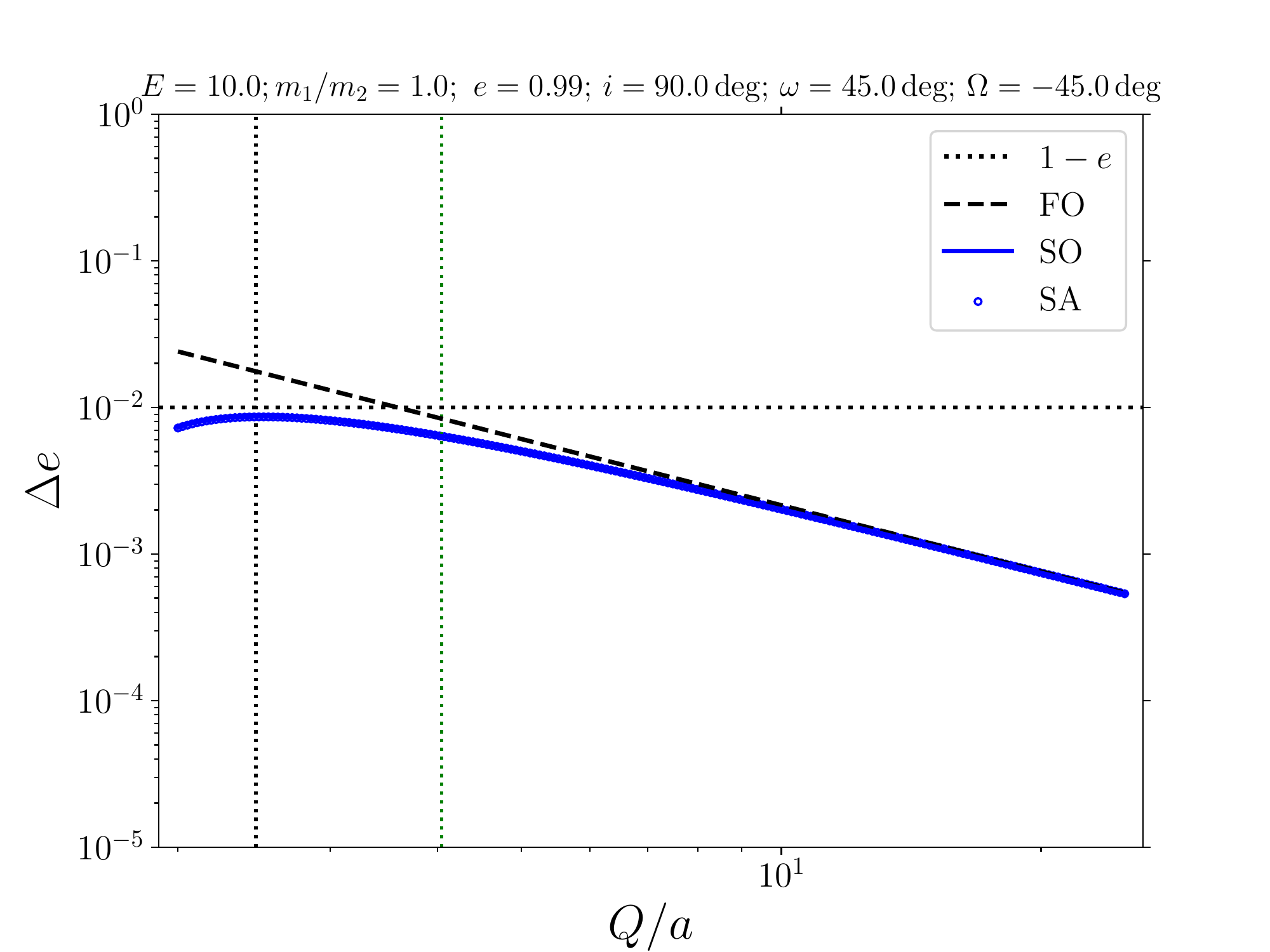}
\caption { Scalar eccentricity changes as a function of $Q/a$ for different values of $\eper$ and $e$ (indicated in the top of each panel). Refer to the caption of \F~\ref{fig:series1} for a description of the meaning of the various lines and symbols. }
\label{fig:series2}
\end{figure}

In \F~\ref{fig:series2}, we show similar results for different parameter choices in four panels with $\eper=1$ (parabolic encounter) or $\eper=10$, and $e=0.99$ or $e=0.999$. The SO expression generally gives a good description. Some deviation can be seen in the case when $\eper=1$ and $e=0.99$, which can be attributed to higher-order terms in $\epssa$ (i.e., $\epssa^3$) that are important for small $Q/a$. Higher-order terms are left for future work.

\subsubsection{Useful properties of the SO solution}
\label{sect:ex:series:prop}
Evidently, the SO solution removes the need for numerical integration of the equations of motion. The latter can be achieved with the SA approximation, or with the most general but also most computationally expensive method of direct 3-body integration. The implied speed-up can be advantageous when a large number of encounters need to be evaluated, e.g., in Monte Carlo approaches. 

In addition, the SO equations can be used to gain more insight into the effect of secular encounters. In the case of the series of $Q/a$ discussed in the previous section, the value of $\epssa$ (corresponding to a particular $Q/a$) for the sign change of $\Delta e$ can be simply expressed using equation~(\ref{eq:Delta_e_SO}), i.e., 
\begin{align}
\label{eq:epssade}
\epssade = - \frac{\unit{e} \cdot \ve{f}_e}{\unit{e} \cdot \ve{g}_e}.
\end{align}
The value of $\epssa$ corresponding to the plateau ($\partial \Delta e/\partial (Q/a) = 0$, or, equivalently, $\partial \Delta e/\partial \epssa=0$) is easily found to be
\begin{align}
\label{eq:epssaplat}
\epssaplat =- \frac{1}{2} \frac{\unit{e} \cdot \ve{f}_e}{\unit{e} \cdot \ve{g}_e} = \frac{1}{2} \epssade.
\end{align}
In terms of $Q/a$ (see equation~\ref{eq:epssa}), the latter implies 
\begin{align}
\label{eq:Q_div_a_relation}
(Q/a)_{\mathrm{plateau}} = 2^{2/3} (Q/a)_{\Delta e=0} \simeq 1.59 \, (Q/a)_{\Delta e=0}.
\end{align}
In Figs~\ref{fig:series1} and \ref{fig:series2}, we show in each $\Delta e$-panel with the two vertical black dotted lines the values of $Q/a$ corresponding to the sign change and plateau according to equations~(\ref{eq:epssade}) and (\ref{eq:epssaplat}).

Although the expression in equation~(\ref{eq:epssade}) seems simple, the explicit relation in terms of $\eper$, $\ve{e}$ and $\ve{\j}$ is very complicated (this can be mainly attributed to the complicated function $\geii$). Fortunately, in the limit of parabolic orbits, $\ve{g}_e$ greatly simplifies (see \S~\ref{sect:cor:calc:par}); in this case, $\epssade$ reduces to
\begin{align}
\label{eq:epssade_par}
\nonumber \nonumber \epssade &= -640 \sqrt{1-e^2} \sin 2\omega \sin^2i \\
\nonumber &\quad \times \Biggl [100 \left(1-e^2\right) \sin 2 \omega  \biggl \{ \left (5 \cos i+3 \cos 3 i\right ) \cos 2 \Omega +6 \sin i \sin 2 i\biggl \}+4 \cos 2 i \biggl \{3 \pi  \left(81 e^2-56\right)+200 \left(1-e^2\right) \cos 2 \omega  \sin 2 \Omega \biggl \} \\
&\qquad \qquad +3 \pi  \biggl \{ 200 e^2 \sin ^4i \cos 4 \omega +8 \left(16 e^2+9\right) \sin ^2 2 i \cos 2 \omega +\left(39 e^2+36\right) \cos 4 i-299 e^2+124\biggl \}\Biggl ]^{-1}.
\end{align}

Another property that immediately follows from the SO solution is the slope in the small $Q/a$ regime in which $\Delta e$ is negative. In the small $Q/a$ regime, the term $\propto \epssa^2$ in equation~\ref{eq:Delta_e_SO}) dominates. Therefore, $|\Delta e|\propto \epssa^2 \propto (Q/a)^{-3}$, which is consistent with the numerical results.

\section{Discussion}
\label{sect:discussion}
\subsection{Higher-order expansion terms}
\label{sect:discussion:oct}

\begin{figure}
\center
\includegraphics[scale = 0.46, trim = 8mm 0mm 8mm 0mm]{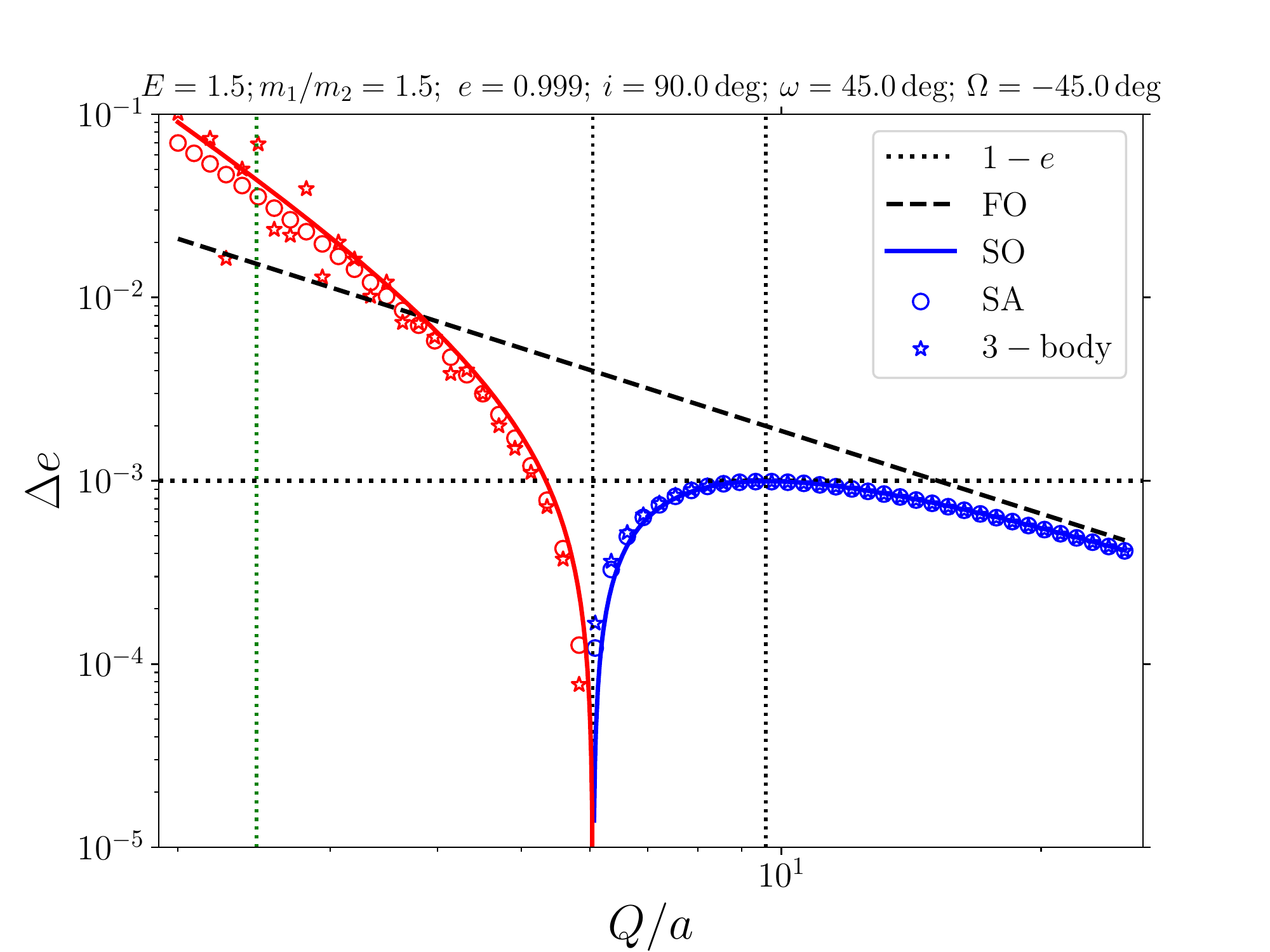}
\includegraphics[scale = 0.46, trim = 8mm 0mm 8mm 0mm]{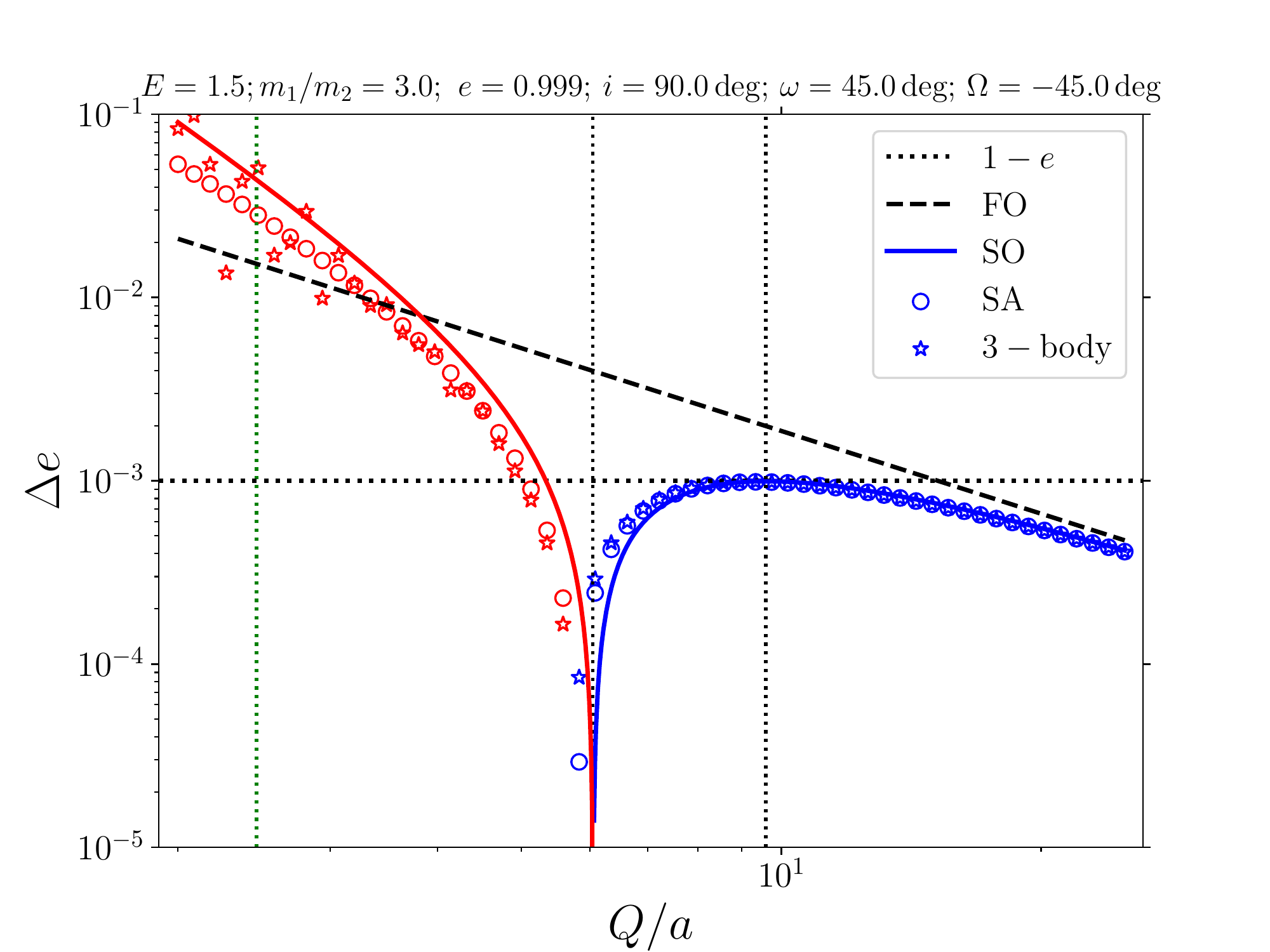}
\caption { Similar to \F~\ref{fig:series1}, here for unequal binary mass cases, and, in the case of the SA relations, with the inclusion of the octupole-order terms (see equations~\ref{eq:EOM_SA_gen_oct}). The `SO' lines include the FO octupole-order prediction, equation~(\ref{eq:Delta_e_FO_oct}). Left-hand panel: $m_1=1.2$ and $m_2=0.8$; right-hand panel: $m_1=1.5$ and $m_2=0.5$. The perturber mass in both cases is $\mper=1$. }
\label{fig:series3}
\end{figure}

Our derivation of the SO corrections were based on the quadrupole-order SA equations of motion, equations~(\ref{eq:EOM_SA_gen}). However, for small $Q/a$, higher-order terms in the Hamiltonian expansion in terms of $x=r/R$ are generally also important. In particular, the next-order terms in the Hamiltonian are the octupole-order terms ($\propto x^3$), which give rise to terms in the SA equations of motion given by (e.g., \citealt{2018MNRAS.476.4139H})
\begin{subequations}
\label{eq:EOM_SA_gen_oct}
\begin{align}
\footnotesize \left ( \frac{\mathrm{d}\ve{e}}{\mathrm{d} \theta}\right )_{\mathrm{oct}} &= \epssa \epsoct (1+\eper \cos \theta)^2 \frac{15}{16} {\footnotesize \Biggl [ 16 \left(\ve{e} \cdot \unit{R} \right ) \left(\ve{\j}\times \ve{e} \right ) - \left(1-8e^2 \right) \left(\ve{\j} \times \unit{R} \right ) + 10 \left ( \ve{e} \cdot \unit{R} \right ) \left ( \ve{\j} \cdot \unit{R} \right ) \left ( \ve{e} \times \unit{R} \right ) + 5 \left(\ve{\j}\cdot \unit{R} \right )^2 \left ( \ve{\j}\times \unit{R} \right ) - 35 \left ( \ve{e} \cdot \unit{R} \right )^2 \left ( \ve{\j} \times \unit{R} \right ) \Biggl ];} \\
\left ( \frac{\mathrm{d}\ve{\j}}{\mathrm{d} \theta} \right)_{\mathrm{oct}}  &= \epssa \epsoct (1+\eper \cos \theta)^2 \frac{15}{16} \Biggl [ - \left(1-8e^2 \right) \left(\ve{e} \times \unit{R} \right ) + 10 \left ( \ve{e} \cdot \unit{R} \right ) \left ( \ve{\j} \cdot \unit{R} \right ) \left ( \ve{\j} \times \unit{R} \right ) + 5 \left(\ve{\j}\cdot \unit{R} \right )^2 \left ( \ve{e}\times \unit{R} \right ) - 35 \left ( \ve{e} \cdot \unit{R} \right )^2 \left ( \ve{e} \times \unit{R} \right ) \Biggl ].
\end{align}
\end{subequations}
Here, the `octupole parameter' $\epsoct$ (analogous to the octupole parameter in hierarchical triples, e.g., \citealt{2011ApJ...742...94L,2011PhRvL.107r1101K,2013ApJ...779..166T,2014ApJ...791...86L}, or more generally in hierarchical systems, \citealt{2016MNRAS.459.2827H}) is defined by
\begin{align}
\label{eq:epsoct}
\epsoct \equiv \frac{|m_1-m_2|}{m_1+m_2} \frac{a}{Q}  \frac{1}{1+\eper}.
\end{align}

In the examples in \S~\ref{sect:ex}, we chose $m_1=m_2$ implying that the octupole-order terms vanish; the quadrupole-order SO results generally agreed with the 3-body integrations. However, for unequal masses of the binary components, the octupole-order terms do not vanish, and can give significant contributions for small $Q/a$. This is illustrated in \F~\ref{fig:series3}, in which, similarly to \F~\ref{fig:series1} but here for unequal binary mass component cases, $\Delta e$ is plotted as a function of $Q/a$. In the case of the SA integrations, the octupole-order terms (equations~\ref{eq:EOM_SA_gen_oct}) were included. The `SO' lines include the FO octupole-order prediction, which can be obtained by integrating equations~(\ref{eq:EOM_SA_gen_oct}) assuming constant $\ve{e}$ and $\ve{\j}$, yielding a scalar eccentricity change given by
\begin{align}
\label{eq:Delta_e_FO_oct}
\nonumber (\Delta e)_{\mathrm{FO,oct}} &\simeq \frac{5}{32 \eper^2 e} \epssa \epsoct \Biggl [ 3 \eper^3 L \left(e_x^2 (3 e_y \j_z-73 e_z \j_y)+10 e_x \j_x (7 e_y e_z+\j_y \j_z)+e_y \j_z \left(3 e_y^2-32 e_z^2-15 \j_x^2-5 \j_y^2+4\right) \right. \\
\nonumber &\quad \left. +e_z \j_y \left(-3 e_y^2+32 e_z^2+5 \j_x^2+5 \j_y^2-4\right)\right)+\sqrt{1-\frac{1}{\eper^2}} \left(e_x^2 \left(3 \left(16 \eper^4-27 \eper^2+14\right) e_y \j_z-\left(160 \eper^4+45 \eper^2+14\right) e_z \j_y\right) \right. \\
\nonumber &\quad \left. +2 \left(8 \eper^4+9 \eper^2-2\right) e_x \j_x (7 e_y e_z+\j_y \j_z)-e_y \j_z \left(8 \eper^4 \left(e_y^2+8 e_z^2+4 \j_x^2+\j_y^2-1\right)+\eper^2 \left(-31 e_y^2+32 e_z^2+11 \j_x^2+9 \j_y^2-4\right) \right. \right. \\
&\quad \left. \left.+2 \left(7 e_y^2+\j_x^2-\j_y^2\right)\right)+e_z \j_y \left(8 \eper^4 \left(e_y^2+8 e_z^2+2 \j_x^2+\j_y^2-1\right)+\eper^2 \left(-31 e_y^2+32 e_z^2-7 \j_x^2+9 \j_y^2-4\right)+14 e_y^2+6 \j_x^2-2 \j_y^2\right)\right) \Biggl ].
\end{align}
In the left-hand panel, we set $m_1=1.2$ and $m_2=0.8$; in the right-hand panel, $m_1=1.5$ and $m_2=0.5$. The perturber mass in both cases is $\mper=1$. 

\F~\ref{fig:series3} shows that our SO prediction with the inclusion of the FO octupole-order terms does not accurately describe $\Delta e$, particularly as the difference $|m_1-m_2|$ increases. The octupole-order SA equations still give good agreement with the 3-body integrations, provided that $Q/a\gtrsim 3$ (otherwise, the SA approximation breaks down). 

This discrepancy can be understood by noting that the octupole-order equations of motion give rise to additional terms in the Fourier expansion of $\ve{e}$ and $\ve{\j}$, equations~(\ref{eq:ej_four}), of order $\epssa \epsoct$ (and higher orders). Consequently, the lowest-order contribution from the SO octupole-order term will be of the order of $\epssa^2 \epsoct$ (and not $\epssa^2 \epsoct^2$, as might be naively expected). 

A derivation of the SO octupole-order terms is beyond the scope of this paper, and is left for future work.

\subsection{Third body's orbit}
\label{sect:discussion:orbit}
In our calculations, we assumed that the perturber's orbit is perfectly hyperbolic or parabolic. In reality, as the perturber's orbit becomes less secular (in particular, for smaller $Q/a$), deviations from hyperbolic or parabolic orbits may become increasingly important. Analytic calculations taking into account such deviations are beyond the scope of this work. However, such corrections may not even be useful, since the SA approximation typically breaks down in similar regimes as $\mathcal{R}$ (equation~\ref{eq:R_def}) approaches unity. 

\begin{figure}
\center
\includegraphics[scale = 0.47, trim = 8mm 0mm 8mm 0mm]{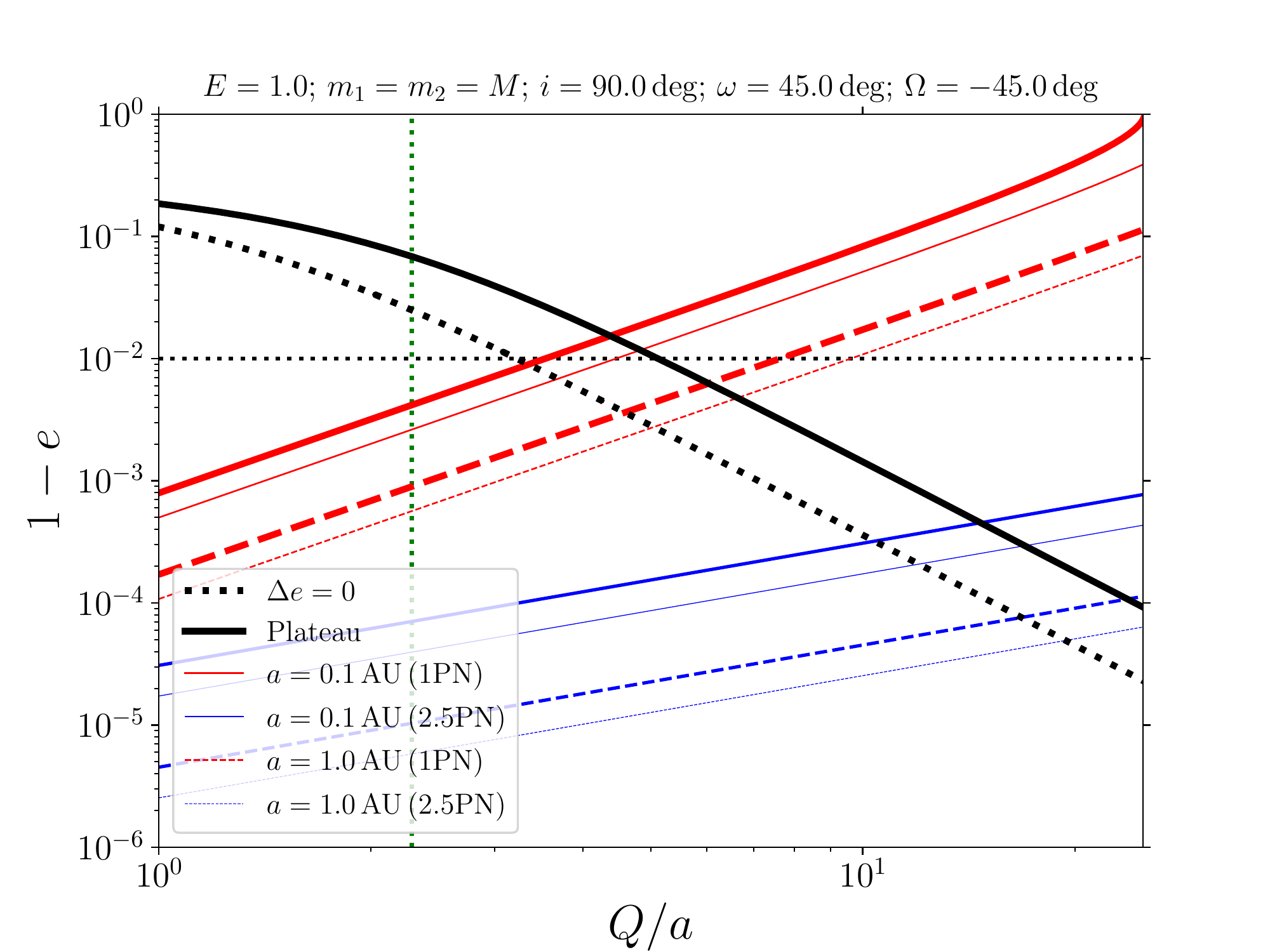}
\includegraphics[scale = 0.47, trim = 8mm 0mm 8mm 0mm]{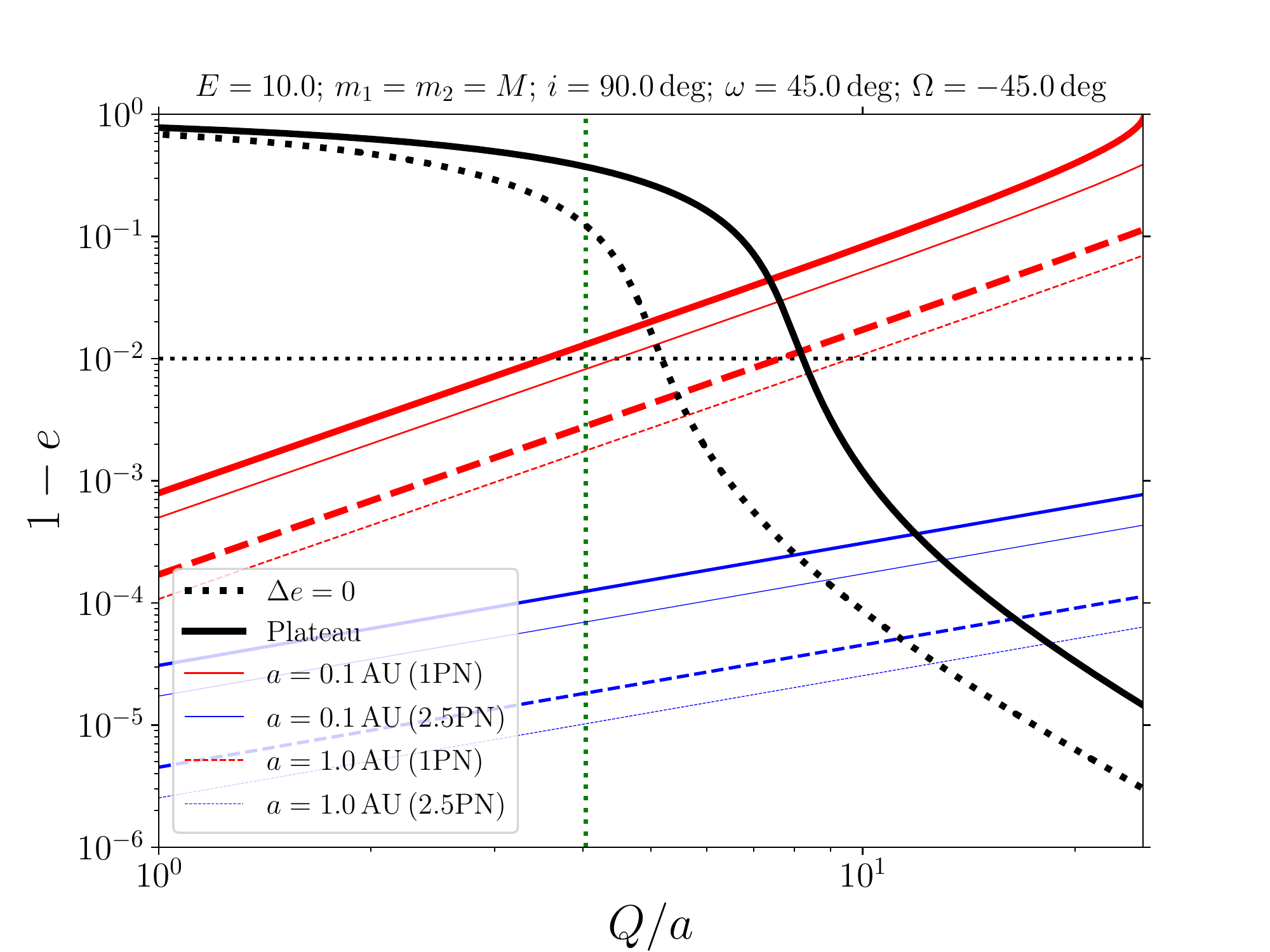}
\caption {Locations in the $(Q/a,1-e)$ parameter space (where $e$ is the initial binary eccentricity) for which the sign change in $\Delta e$ occurs (dotted black lines), and for which $\Delta e$ reaches the plateau (solid black lines), according to our SO results (see \S~\ref{sect:cor:calc}). The masses are $m_1=m_2=\mper$, and the (fixed) orbital binary angles angles are $i=90^\circ$, $\omega=45^\circ$, $\Omega=-45^\circ$; the left (right)-hand panels correspond to $\eper=1$ ($\eper=10$). Red lines show equation~(\ref{eq:e_1PN}), i.e., the lines below which 1PN terms are expected to be important during the encounter, where solid (dashed) lines correspond to $a=0.1\,\au$ ($a=1\,\au)$, and thick (thin) lines correspond to $m_1=m_2=\mper = 20\,\msun$ ($m_1=m_2=\mper=10\,\msun$). Blue lines show equation~(\ref{eq:e_25PN}), i.e., lines below which 2.5PN terms are expected to be important during the encounter. }
\label{fig:overview}
\end{figure}

\subsection{Applicability to star clusters}
\label{sect:discussion:appl}
We assumed that the (dynamical) effects of other stars on the evolution of a binary in a star cluster can be described in terms of a restricted problem in which a single perturber approaches the binary on a parabolic or hyperbolic orbit. Reality is, of course, more complicated. For example, due to perturbations from other stars, the orbit of the third body relative to the binary is never exactly parabolic or hyperbolic. However, the effect of the third body on the binary is highly concentrated near periapsis, since $\mathrm{d}e/\mathrm{d} t \propto R^{-3}$. Therefore, we expect that the exact shape of the orbit at larger distances from the binary is not very important for the evolution of the binary. Also, in a real system, the cluster potential (i.e., other stars viewed as a whole) can affect the binary, depending on the potential properties (see, e.g., \citealt{2019arXiv190201344H,2019arXiv190201345H}). 

Another related point of concern is that not all encounters are safely in the secular regime, which we assumed here. The fraction of secular encounters can be crudely estimated as $1-(b_\mathrm{crit}^2/b_\mathrm{max}^2)$, where $b_\mathrm{crit}$ is the critical impact parameter delineating secular from nonsecular encounters\footnote{Of course, in reality there is no sharp boundary between secular and non-secular encounters.}, and $b_\mathrm{max}$ is the largest impact parameter considered. For the typical parameters assumed in this paper, $b_\mathrm{crit} \sim 2 \, a$ (see, e.g., the vertical green dotted lines in Figs~\ref{fig:series1}, \ref{fig:series2}, and \ref{fig:series3}), implying a fraction of secular encounters for given $b_\mathrm{max}$ given by $\sim 1 - 0.04 \, [b_\mathrm{max}/(10 a)]^{-2}$. If $b_\mathrm{max}$ is taken to be $0.1\, \mathrm{pc} \approx 2\times 10^4\,\au$, a tenth of the typical core radius of a globular cluster and roughly the typical size of dense subclusters of BHs formed in globular clusters (e.g., \citealt{2010MNRAS.402..371B}), then the fraction of secular encounters for a binary with a semimajor axis of $1\,\au$ is $\sim 1-10^{-8}$. Clearly, most encounters are secular. Evidently, although non-secular encounters are rare, they can dramatically affect the binary. The combined effect of strong and weak encounters is discussed in \citet{2019arXiv190607189S}.

\subsection{Overview of parameter space}
\label{sect:discussion:overview}
We showed that the (scalar) eccentricity change due to secular encounters can significantly deviate from the analytic prediction of \citet{1996MNRAS.282.1064H} when the initial binary eccentricity is high. Using our analytic SO expressions, we here quantify the parameter space in which this deviation is important, focusing on BH binaries in dense stellar clusters and also considering the importance of post-Newtonian (PN) terms.

In \F~\ref{fig:overview}, we show, in the $(Q/a,1-e)$ parameter space (where $e$ is the initial binary eccentricity), locations for which the sign change in $\Delta e$ occurs (dotted black lines), and for which $\Delta e$ reaches a plateau (solid black lines). Generally, these two values are given by equations~(\ref{eq:epssade}) and (\ref{eq:epssaplat}), respectively (also note that there exists a general relation between the two as described by equation~\ref{eq:Q_div_a_relation}). Here, we set $m_1=m_2=\mper$, $i=90^\circ$, $\omega=45^\circ$, $\Omega=-45^\circ$; the left (right)-hand panels correspond to $\eper=1$ ($\eper=10$). For a given value of $Q/a$, the black lines indicate the initial eccentricity for which the sign change and plateau phenomena occur; above the black dashed line (smaller eccentricity), the result of \citeauthor{1996MNRAS.282.1064H} (\citeyear{1996MNRAS.282.1064H}; i.e., first order in $\epssa$) is expected to give an at least reasonable description of $\Delta e$, whereas below it, higher-order terms in $\epssa$ (second order and higher) become important. 

Generally, the initial binary eccentricities for which $\Delta e$ reaches a sign change or plateau increase with increasing $Q/a$. A larger initial $e$ implies that a smaller change in the binary eccentricity can push the binary towards unity eccentricity (which is associated with the sign change and plateau phenomena), implying a weaker perturbation and hence a larger value of $Q/a$. In the case of $\eper=1$ and a perturber with $Q/a\sim 4$, a binary needs to have an initial eccentricity of $\sim 0.99$ for deviations from \citet{1996MNRAS.282.1064H} to be important. For $Q/a\sim 10$, the required eccentricity is $\sim 0.999$. From equation~(\ref{eq:epssade_par}), it follows that in the limit of parabolic orbits and high initial binary eccentricity, $\epssade \propto \sqrt{1-e^2}$, i.e., $1-e \propto (Q/a)^{-3}$, consistent with the high-eccentricity behaviour shown with the black lines in the left-hand panel in \F~\ref{fig:overview}.

PN terms are potentially important for BH binaries in dense stellar systems. Here, we estimate the importance of the lowest-order PN terms, which are of the order of $(v/c)^2$, where $v$ is the binary's orbital speed and $c$ is the speed of light. The 1PN terms give rise to precession of the binary's line of apsides at a rate given by (e.g., \citealt{1972gcpa.book.....W})
\begin{align}
\label{eq:1PN}
\dot{\omega}_\oPN = \frac{3}{1-e^2} \frac{\rg}{a} n_\bin.
\end{align}
Here, $\rg\equiv Gm/c^2$ is the binary's gravitational radius, and $n_\bin \equiv \sqrt{Gm/a^3}$ is the binary's mean motion. Analogously to triple systems (e.g., \citealt{2002ApJ...578..775B,2003ApJ...589..605W,2007ApJ...669.1298F,2011ApJ...741...82T,2013ApJ...773..187N,2015MNRAS.447..747L}) and orbits around a massive BH (e.g., \citealt{2011PhRvD..84d4024M,2014MNRAS.437.1259B,2014MNRAS.443..355H,2014CQGra..31x4003B,2016ApJ...820..129B,2018ApJ...860L..23B}), rapid relativistic precession can decrease the efficiency of the Newtonian torque exerted by the third body. To investigate the importance of the 1PN terms, it is therefore relevant to compare the relativistic rate, equation~(\ref{eq:1PN}), to the (Newtonian) precession rate associated with the third body, $\dot{\omega}_\TB$. For simplicity, we evaluate the latter using the quadrupole-order term at periapsis of the perturber ($\theta=0$), giving (see also equation~\ref{eq:EOM_SA_gen})
\begin{align}
\dot{\omega}_\TB(\theta=0) = \frac{1}{e} \, \unit{q} \cdot \frac{\mathrm{d} \ve{e}}{\mathrm{d} t}(\theta=0) = \frac{3}{64} n_\bin \sqrt{1-e^2} \frac{\mper}{m} \left (\frac{a}{Q} \right )^3 f_{\TB,\dot{\omega}}(i,\omega,\Omega).
\end{align}
Here, $\unit{q} \equiv \unit{\ve{\j}} \times \unit{\ve{e}}$, and $f_{\TB,\dot{\omega}}(i,\omega,\Omega)$ depends on the binary orientation and is given by
\begin{align}
\nonumber &f_{\TB,\dot{\omega}}(i,\omega,\Omega) = 4-20 \cos (i+2 \omega -2 \Omega )+5 \cos (2 (i-\omega -\Omega ))+5 \cos (2 (i+\omega -\Omega ))+5 \cos (2 (i-\omega +\Omega ))+5 \cos (2 (i+\omega +\Omega ))\\
\nonumber &\quad -20 \cos (i-2 \omega +2 \Omega )+20 \cos (i-2 (\omega +\Omega ))+20 \cos (i+2 (\omega +\Omega ))-10 \cos (2 (i-\omega ))-10 \cos (2 (i+\omega ))\\
&\quad -6 \cos (2 (i-\Omega ))-6 \cos (2 (i+\Omega ))+12 \cos (2 i)+30 \cos (2 (\omega -\Omega ))+30 \cos (2 (\omega +\Omega ))+20 \cos 2 \omega +12 \cos 2 \Omega .
\end{align}
Setting $|\dot{\omega}_\oPN| = |\dot{\omega}_\TB(\theta=0)|$ gives the following relation for the binary eccentricity above which we expect 1PN terms to become important,
\begin{align}
\label{eq:e_1PN}
\left( 1-e_\oPN^2\right)^{3/2} = \frac{64}{|f_{\TB,\dot{\omega}}(i,\omega,\Omega)|} \frac{m}{\mper} \frac{\rg}{a} \left(\frac{Q}{a} \right )^3.
\end{align}

Equation~(\ref{eq:e_1PN}) is shown in \F~\ref{fig:overview} with the red lines, where solid (dashed) lines correspond to $a=0.1\,\au$ ($a=1\,\au)$, and thick (thin) lines correspond to $m_1=m_2=\mper = 20\,\msun$ ($m_1=m_2=\mper=10\,\msun$). Note that the 1PN terms break the scale-invariance of the problem, hence explicit values for the masses and the binary semimajor axis should be specified. For a given $Q/a$, 1PN terms are important for any initial binary eccentricity with $e>e_\oPN$ (points below the red lines). For high eccentricities, equation~(\ref{eq:e_1PN}) gives the scaling $1-e_\oPN \propto (Q/a)^2$, consistent with \F~\ref{fig:overview}. The red lines in \F~\ref{fig:overview} show that 1PN terms are potentially important for encounters with highly eccentric and tight binaries. 

A higher order PN effect is the dissipation of orbital energy and angular momentum by GWs. The (lowest-order) associated PN term, the 2.5PN term, gives rise to an eccentricity decay rate given by \citep{1964PhRv..136.1224P}
\begin{align}
\label{eq:2.5PN}
\dot{e}_\tPN = - \frac{304}{15} e \frac{G^3 m_1m_2m}{c^5 a^4(1-e^2)^{5/2}} \left(1+ \frac{121}{304}e^2 \right ) \approx - \frac{85}{3} e \frac{G^3 m_1m_2m}{c^5 a^4(1-e^2)^{5/2}},
\end{align}
where after the approximation sign we assumed $e\rightarrow 1$. The eccentricity change due to the third body, evaluated at $\theta=0$ and at the quadrupole order, is given by (see also equation~\ref{eq:EOM_SA_gen})
\begin{align}
\dot{e}_\TB(\theta=0) = \unit{e} \cdot \frac{\mathrm{d} \ve{e}}{\mathrm{d} t} (\theta=0) = \frac{15}{16} n_\bin e \sqrt{1-e^2} \frac{\mper}{m} \left ( \frac{a}{Q} \right )^3 f_{\TB,e}(i,\omega,\Omega),
\end{align}
where
\begin{align}
f_{\TB,e}(i,\omega,\Omega) = \sin2 \omega \left [ (3+\cos2i) \cos 2\Omega+2\sin^2i \right] + 4 \cos i \cos2\omega \sin2\Omega.
\end{align}
The dissipative PN terms become important {\it during} the passage of the perturber if $|\dot{e}_\tPN| = |\dot{e}_\TB(\theta=0)|$, i.e., when (in the limit of high binary eccentricity)
\begin{align}
\label{eq:e_25PN}
\left(1-e_\tPN^2\right)^3 = \frac{272}{9|f_{\TB,e}(i,\omega,\Omega)|} \frac{m_1m_2}{m \mper} \left ( \frac{\rg}{a} \right )^{5/2} \left(\frac{Q}{a} \right )^3.
\end{align}
Equation~(\ref{eq:e_25PN}) is shown in \F~\ref{fig:overview} with the blue lines, where solid (dashed) lines correspond to $a=0.1\,\au$ ($a=1\,\au)$, and thick (thin) lines correspond to $m_1=m_2=\mper = 20\,\msun$ ($m_1=m_2=\mper=10\,\msun$). The scaling inferred from equation~(\ref{eq:e_25PN}) is $1-e_\tPN \propto (Q/a)$, consistent with \F~\ref{fig:overview}. 

Clearly, the dissipative PN terms are only important for binaries with extremely high eccentricity. However, such binaries are likely to merge rapidly due to GW emission before any encounter can change the binary's properties. The latter is illustrated in \F~\ref{fig:overview} with the horizontal black dotted line, which shows the typical eccentricity for which, in a globular cluster, a BH binary is expected to merge due to GW emission before a next encounter (e.g., \citealt{2018MNRAS.481.5445S}).

A caveat of the criterion equation~(\ref{eq:e_25PN}) is that the dissipative term in equation~(\ref{eq:2.5PN}) is evaluated at the {\it initial} binary eccentricity. In reality, the binary eccentricity changes during the encounter, potentially significantly reducing the merger time-scale (this issue will be addressed in future work, \citealt{2019arXiv190607189S}). A complete calculation of the SO terms with the inclusion of PN terms is beyond the scope of this paper.

\section{Conclusions}
\label{sect:conclusions}
We presented an analytic calculation of the secular effect on a binary of a perturber moving on a hyperbolic or parabolic orbit. We extended previous works \citep{1996MNRAS.282.1064H,2018MNRAS.476.4139H} in which first-order (FO) perturbation theory was used, by applying second-order (SO) perturbation techniques. Our results can be used to evaluate the importance of distant encounters on the evolution of eccentric binaries in dense stellar systems. The conclusions are given below.

\medskip \noindent 1. In the FO approximation, the binary's orbital elements are assumed to be constant during the passage of the third object. However, in some cases, particularly when the binary eccentricity is high (specifically, if $\Delta e \sim 1-e$), the FO approximation can break down due to a relatively large change of the binary's orbital elements during the passage. We used a Fourier series expansion to derive SO correction terms, which take into account the binary's varying orbital elements. We derived explicit expressions for the change in the eccentricity and angular-momentum vectors of the binary, $\Delta \ve{e}$ and $\Delta \ve{\j}$, respectively, as a function of the initial parameters of the system. 

\medskip \noindent 2. The general SO expressions for $\Delta \ve{e}$ and $\Delta \ve{\j}$ are very complicated. We provided a freely-available and simple \textsc{Python} script to evaluate them, including, for reference and testing purposes, routines to numerically integrate the singly-averaged (SA) equations of motion, and the non-averaged, direct 3-body equations of motion (see \S~\ref{sect:ex} for the url). In the limit of parabolic encounters ($\eper\rightarrow1$), the SO expressions become manageable, leading to a simple closed-form expression for the scalar eccentricity change (equation~\ref{eq:Delta_e_CDA_par}). The latter can be used to quickly explore the importance of the SO term (without a need to use our supplied \textsc{Python} script). 

\medskip \noindent 3. We tested the SO expressions for a number of cases with highly eccentric binaries. Such binaries are formed as a result of strong dynamical encounters in dense stellar systems. We showed that, as the perturber periapsis distance $Q$ decreases, $\Delta e$ reaches a plateau, and subsequently drops and changes sign. For even smaller $Q$, $|\Delta e|$ increases with decreasing $Q$ with a steeper slope compared to larger $Q$. We showed that our SO expressions accurately describe this phenomenon, in the limit when the octupole-order terms are unimportant (e.g., if $m_1=m_2$), and the secular approximation applies (i.e., it is justified to average over the binary's motion). In contrast, the (quadrupole-order) FO expression from \citet{1996MNRAS.282.1064H} predicts incorrect and unphysical eccentricities in this regime (in some cases, even $e+\Delta e > 1$). 

\medskip \noindent 4. Our SO expressions were derived from the quadrupole-order equations of motion (i.e., obtained by an expansion of the Hamiltonian to second order in the separation of the binary to the separation of the third body relative to the binary's center of mass). For small perturber periapsis distances, higher-order terms can be important. We briefly discussed these terms in \S~\ref{sect:discussion:oct}, but a computation of the associated SO terms is left for future work.

\medskip \noindent 5. We briefly explored the parameter space in which the plateau and sign change phenomena described above are important (see \S~\ref{sect:discussion:overview}). We also showed that PN terms are potentially important for reducing the effect of secular encounters for highly eccentric (BH) binaries.

\section*{Acknowledgements}
We thank Scott Tremaine for helpful discussion, and comments on the manuscript. We also thank Stephen McMillan and Piet Hut for stimulating discussions, Douglas Heggie, Fred Rasio, Evgeni Grishin, and Chris Hamilton for comments on the manuscript, and the anonymous referee for a helpful report. A.S.H. gratefully acknowledges support from the Institute for Advanced Study, and the Martin A. and Helen Chooljian Membership. J.S. acknowledges support from the Lyman Spitzer Fellowship.

\bibliographystyle{mnras}
\bibliography{literature}

\appendix

%\onecolumn

\section{Explicit expressions}
\label{app}

\subsection{Fourier coefficients}
\label{app:fourier}
Here, we give the explicit expressions for the Fourier coefficients of the equations of motion (with $l\neq0$), defined in equations~(\ref{eq:gen_fourier}).
\begin{subequations}
\begin{align}
\nonumber \etcl &= \frac{3}{\pi  \eper^2 l f_l} \Leper \Biggl [ \pi ^5 \sqrt{1-\frac{1}{\eper^2}} \eper l^5 \cos (\pi  l) \left(\left(3 \eper^2-5\right) e_y \j_z+\left(\eper^2+1\right) e_z \j_y\right)-\pi ^4 l^4 \Leper \sin (\pi  l) \left(\left(25-23 \eper^2\right) e_y \j_z+\left(3 \eper^2-5\right) e_z \j_y\right)\\
\nonumber &\quad-\pi ^3 \sqrt{1-\frac{1}{\eper^2}} \eper l^3 \Leper^2 \cos (\pi  l) \left(\left(9 \eper^2+25\right) e_y \j_z+\left(19 \eper^2-5\right) e_z \j_y\right) -\pi ^2 l^2 \Leper^3 \sin (\pi  l) \left(\left(59 \eper^2-25\right) e_y \j_z+\left(9 \eper^2+5\right) e_z \j_y\right)\\
\nonumber &\quad+6 \pi  \sqrt{1-\frac{1}{\eper^2}} \eper l \Leper^4 \cos (\pi  l) \left(\left(5-2 \eper^2\right) e_y \j_z+\left(10 \eper^2-1\right) e_z \j_y\right)+18 \eper^2 \Leper^5 \sin (\pi  l) (e_y \j_z+3 e_z \j_y)\Biggl ] \,\unit{x} \\
\nonumber &\quad -\frac{3}{\pi  \eper^2 l f_l} \Leper \Biggl [ \pi ^5 \sqrt{1-\frac{1}{\eper^2}} \eper l^5 \cos (\pi  l) \left(\left(5-2 \eper^2\right) e_x \j_z+\left(2 \eper^2-1\right) e_z \j_x\right)+\pi ^4 l^4 \Leper \sin (\pi  l) \left(\left(25-22 \eper^2\right) e_x \j_z+\left(6 \eper^2-5\right) e_z \j_x\right) \\
\nonumber &\quad -\pi ^3 \sqrt{1-\frac{1}{\eper^2}} \eper l^3 \Leper^2 \cos (\pi  l) \left(\left(4 \eper^2-25\right) e_x \j_z+5 e_z \left(4 \eper^2 \j_x+\j_x\right)\right)-\pi ^2 l^2 \Leper^3 \sin (\pi  l) \left(\left(25-46 \eper^2\right) e_x \j_z+5 \left(6 \eper^2-1\right) e_z \j_x\right) \\
\nonumber &\quad+6 \pi  \sqrt{1-\frac{1}{\eper^2}} \eper l \Leper^4 \cos (\pi  l) \left(\left(8 \eper^2-5\right) e_x \j_z+e_z \left(8 \eper^2 \j_x+\j_x\right)\right)+18 \eper^2 \Leper^5 \sin (\pi  l) (e_x \j_z+3 e_z \j_x)\Biggl ]\,\unit{y} \\
\nonumber &\quad -\frac{3}{\pi  \eper^2 l f_l} \Leper \Biggl [\pi ^5 \sqrt{1-\frac{1}{\eper^2}} \eper l^5 \cos (\pi  l) \left(\left(\eper^2-4\right) e_x \j_y+\left(3 \eper^2-4\right) e_y \j_x\right)+\pi ^4 l^4 \Leper \sin (\pi  l) \left(\left(17 \eper^2-20\right) e_x \j_y+\left(19 \eper^2-20\right) e_y \j_x\right)\\
\nonumber &\quad-\pi ^3 \sqrt{1-\frac{1}{\eper^2}} \eper l^3 \Leper^2 \cos (\pi  l) \left(\left(20-11 \eper^2\right) e_x \j_y+5 \left(3 \eper^2+4\right) e_y \j_x\right)-\pi ^2 l^2 \Leper^3 \sin (\pi  l) \left(\left(29 \eper^2-20\right) e_x \j_y+5 \left(11 \eper^2-4\right) e_y \j_x\right) \\
&\quad+12 \pi  \sqrt{1-\frac{1}{\eper^2}} \eper l \Leper^4 \cos (\pi  l) \left(\left(2-5 \eper^2\right) e_x \j_y+\left(\eper^2+2\right) e_y \j_x\right)+36 \eper^2 \Leper^5 \sin (\pi  l) (e_y \j_x-e_x \j_y)\Biggl ]\,\unit{z}; \\
\nonumber \etsl &= \frac{3}{\eper^2 f_l} \Leper \Biggl [\pi ^4 \left(\eper^2-1\right) l^4 \sin (\pi  l)+\pi ^3 \sqrt{1-\frac{1}{\eper^2}} \eper \left(2 \eper^2-5\right) l^3 \Leper \cos (\pi  l)+\pi ^2 \left(2 \eper^2-5\right) l^2 \Leper^2 \sin (\pi  l)\\
\nonumber &\quad-\pi  \sqrt{1-\frac{1}{\eper^2}} \eper \left(8 \eper^2-5\right) l \Leper^3 \cos (\pi  l)-3 \left(3 \eper^2-2\right) \Leper^4 \sin (\pi  l)\Biggl ] (e_z \j_x-5 e_x \j_z) \, \unit{x} \\
\nonumber &\quad + \frac{3}{\eper^2 f_l} \Leper \Biggl [\pi ^4 \left(-\left(\eper^2-1\right)\right) l^4 \sin (\pi  l)-\pi ^3 \sqrt{1-\frac{1}{\eper^2}} \eper \left(2 \eper^2-5\right) l^3 \Leper \cos (\pi  l)-\pi ^2 \left(2 \eper^2-5\right) l^2 \Leper^2 \sin (\pi  l)\\
\nonumber &\quad+\pi  \sqrt{1-\frac{1}{\eper^2}} \eper \left(8 \eper^2-5\right) l \Leper^3 \cos (\pi  l)+3 \left(3 \eper^2-2\right) \Leper^4 \sin (\pi  l)\Biggl ] (e_z \j_y-5 e_y \j_z)\,\unit{y} \\
\nonumber &\quad + \frac{12}{\eper^2 f_l} \Leper \Biggl [\pi ^4 \left(\eper^2-1\right) l^4 \sin (\pi  l)+\pi ^3 \sqrt{1-\frac{1}{\eper^2}} \eper \left(2 \eper^2-5\right) l^3 \Leper \cos (\pi  l)+\pi ^2 \left(2 \eper^2-5\right) l^2 \Leper^2 \sin (\pi  l)\\
&\quad-\pi  \sqrt{1-\frac{1}{\eper^2}} \eper \left(8 \eper^2-5\right) l \Leper^3 \cos (\pi  l)-3 \left(3 \eper^2-2\right) \Leper^4 \sin (\pi  l)\Biggl ] (e_x \j_x-e_y \j_y)\,\unit{z}; \\
\nonumber \jtcl &= \frac{3}{\pi  \eper^2 l f_l} \Leper \Biggl [\pi ^5 \sqrt{1-\frac{1}{\eper^2}} \eper \left(\eper^2-1\right) l^5 \cos (\pi  l)+5 \pi ^4 \left(\eper^2-1\right) l^4 \Leper \sin (\pi  l)-\pi ^3 \sqrt{1-\frac{1}{\eper^2}} \eper \left(7 \eper^2+5\right) l^3 \Leper^2 \cos (\pi  l)\\
\nonumber &\quad -\pi ^2 \left(17 \eper^2-5\right) l^2 \Leper^3 \sin (\pi  l)+6 \pi  \sqrt{1-\frac{1}{\eper^2}} \eper \left(2 \eper^2+1\right) l \Leper^4 \cos (\pi  l)+18 \eper^2 \Leper^5 \sin (\pi  l)\Biggl ] (5 e_y e_z-\j_y \j_z)\,\unit{x} \\
\nonumber &\quad -\frac{3}{\pi  \eper^2 l f_l} \Leper \Biggl [\pi ^5 \sqrt{1-\frac{1}{\eper^2}} \eper l^5 \cos (\pi  l)-\pi ^4 \left(4 \eper^2-5\right) l^4 \Leper \sin (\pi  l)-\pi ^3 \sqrt{1-\frac{1}{\eper^2}} \eper \left(6 \eper^2-5\right) l^3 \Leper^2 \cos (\pi  l)+\pi ^2 \left(4 \eper^2-5\right) l^2 \Leper^3 \sin (\pi  l)\\
\nonumber &\quad +6 \pi  \sqrt{1-\frac{1}{\eper^2}} \eper \left(4 \eper^2-1\right) l \Leper^4 \cos (\pi  l)+18 \eper^2 \Leper^5 \sin (\pi  l)\Biggl ] (5 e_x e_z-\j_x \j_z)\,\unit{y} \\
\nonumber &\quad + \frac{3}{\eper^2 f_l} \Leper \Biggl [\pi ^4 \left(-\sqrt{1-\frac{1}{\eper^2}}\right) \eper \left(\eper^2-2\right) l^4 \cos (\pi  l)-\pi ^3 \left(9 \eper^2-10\right) l^3 \Leper \sin (\pi  l)+\pi ^2 \sqrt{1-\frac{1}{\eper^2}} \eper \left(\eper^2+10\right) l^2 \Leper^2 \cos (\pi  l)\\
&\quad +12 \sqrt{1-\frac{1}{\eper^2}} \eper \left(\eper^2-1\right) \Leper^4 \cos (\pi  l)+\pi  \left(21 \eper^2-10\right) l \Leper^3 \sin (\pi  l)\Biggl ] (5 e_x e_y-\j_x \j_y)\,\unit{z}; \\
\nonumber \jtsl &= \frac{3}{\eper^2 f_l} \Leper \Biggl [\pi ^4 \left(-\left(\eper^2-1\right)\right) l^4 \sin (\pi  l)-\pi ^3 \sqrt{1-\frac{1}{\eper^2}} \eper \left(2 \eper^2-5\right) l^3 \Leper \cos (\pi  l)-\pi ^2 \left(2 \eper^2-5\right) l^2 \Leper^2 \sin (\pi  l)\\
\nonumber &\quad +\pi  \sqrt{1-\frac{1}{\eper^2}} \eper \left(8 \eper^2-5\right) l \Leper^3 \cos (\pi  l)+3 \left(3 \eper^2-2\right) \Leper^4 \sin (\pi  l)\Biggl ] (5 e_x e_z-\j_x \j_z)\,\unit{x} \\
\nonumber &\quad \frac{3}{\eper^2 f_l} \Leper \Biggl [\pi ^4 \left(\eper^2-1\right) l^4 \sin (\pi  l)+\pi ^3 \sqrt{1-\frac{1}{\eper^2}} \eper \left(2 \eper^2-5\right) l^3 \Leper \cos (\pi  l)+\pi ^2 \left(2 \eper^2-5\right) l^2 \Leper^2 \sin (\pi  l)\\
\nonumber &\quad-\pi  \sqrt{1-\frac{1}{\eper^2}} \eper \left(8 \eper^2-5\right) l \Leper^3 \cos (\pi  l)-3 \left(3 \eper^2-2\right) \Leper^4 \sin (\pi  l)\Biggl ] (5 e_y e_z-\j_y \j_z) \, \unit{y} \\
\nonumber &\quad + \frac{3}{\eper^2 f_l} \Leper \Biggl [\pi ^4 \left(\eper^2-1\right) l^4 \sin (\pi  l)+\pi ^3 \sqrt{1-\frac{1}{\eper^2}} \eper \left(2 \eper^2-5\right) l^3 \Leper \cos (\pi  l)+\pi ^2 \left(2 \eper^2-5\right) l^2 \Leper^2 \sin (\pi  l)\\
&\quad -\pi  \sqrt{1-\frac{1}{\eper^2}} \eper \left(8 \eper^2-5\right) l \Leper^3 \cos (\pi  l)-3 \left(3 \eper^2-2\right) \Leper^4 \sin (\pi  l)\Biggl ] \left(5 e_x^2-5 e_y^2-\j_x^2+\j_y^2\right)\,\unit{z}.
\end{align}
\end{subequations}

Here, $f_l=f_l(\eper)$  is a function of $l$ and $\eper$, given by
\begin{align}
f_l(\eper) = \pi ^6 l^6-14 \pi ^4 l^4 L^2+49 \pi ^2 l^2L^4-36 L^6.
\end{align}

\subsection{Partial SO expressions}
\label{app:SO}
As mentioned in \S~\ref{sect:cor:calc:gen}, the general (i.e., hyperbolic) expressions for $\ge$ and $\gj$ are excessively complicated, and are not given explicitly here (see \S~\ref{sect:ex} for an url to an easy-to-use script in which the general expressions for the eccentricity changes are implemented). Here, we give a partial hyperbolic result in the form of the functions $\gei$ and $\gji$. As discussed in \S~\ref{sect:cor:calc:gen}, these functions can be obtained from the FO result (i.e., $\fe$ and $\fj$) as described by equation~(\ref{eq:gei_gji}), with the substitutions given by equation~(\ref{eq:ej_sub}). The explicit expressions for these substitutions are given below.

\begin{subequations}
\begin{align}
\nonumber e_x' &= -\frac{3}{\eper f_1 f_2 f_3} \sqrt{1-\frac{1}{\eper^2}} \epssa \Leper^3 \left(\left(8 \eper^2-5\right) (f_2 f_3+f_1 (f_2+f_3)) \Leper^2-\pi ^2 \left(2 \eper^2-5\right) (f_2 f_3+f_1 (9 f_2+4 f_3))\right) (e_z \j_x-5 e_x \j_z) \\
&\quad +\frac{1}{4 \eper} \epssa \left(\sqrt{1-\frac{1}{\eper^2}} \left(\left(2 \eper^2-5\right) e_y \j_z+e_z \left(\j_y-10 \eper^2 \j_y\right)\right)-3 \eper \Leper (e_y \j_z+3 e_z \j_y)\right); \\
\nonumber e_y' &= \frac{3}{\eper f_1 f_2 f_3} \sqrt{1-\frac{1}{\eper^2}} \epssa \Leper^3 \left(\left(8 \eper^2-5\right) (f_2 f_3+f_1 (f_2+f_3)) \Leper^2-\pi ^2 \left(2 \eper^2-5\right) (f_2 f_3+f_1 (9 f_2+4 f_3))\right) (e_z \j_y-5 e_y \j_z) \\
&\quad + \frac{1}{4 \eper} \epssa \left(\sqrt{1-\frac{1}{\eper^2}} \left(\left(8 \eper^2-5\right) e_x \j_z+e_z \left(8 \eper^2 \j_x+\j_x\right)\right)+3 \eper \Leper (e_x \j_z+3 e_z \j_x)\right); \\
\nonumber e_z' &= -\frac{12}{\eper f_1 f_2 f_3} \sqrt{1-\frac{1}{\eper^2}} \epssa \Leper^3 \left(\left(8 \eper^2-5\right) (f_2 f_3+f_1 (f_2+f_3)) \Leper^2-\pi ^2 \left(2 \eper^2-5\right) (f_2 f_3+f_1 (9 f_2+4 f_3))\right) (e_x \j_x-e_y \j_y)\\
&\quad +\frac{1}{2 \eper} \sqrt{1-\frac{1}{\eper^2}} \epssa \left(\left(2-5 \eper^2\right) e_x \j_y+\left(\eper^2+2\right) e_y \j_x\right)+3 \eper \epssa \Leper (e_y \j_x-e_x \j_y); \\
\nonumber \j_x' &= \frac{3}{\eper f_1 f_2 f_3} \sqrt{1-\frac{1}{\eper^2}} \epssa \Leper^3 \left(\left(8 \eper^2-5\right) (f_2 f_3+f_1 (f_2+f_3)) \Leper^2-\pi ^2 \left(2 \eper^2-5\right) (f_2 f_3+f_1 (9 f_2+4 f_3))\right) (5 e_x e_z-\j_x \j_z) \\
&\quad -\frac{1}{4 \eper} \epssa \left(\sqrt{1-\frac{1}{\eper^2}} \left(2 \eper^2+1\right)+3 \eper \Leper\right) (5 e_y e_z-\j_y \j_z); \\
\nonumber \j_y' &= -\frac{3}{\eper f_1 f_2 f_3} \sqrt{1-\frac{1}{\eper^2}} \epssa \Leper^3 \left(\left(8 \eper^2-5\right) (f_2 f_3+f_1 (f_2+f_3)) \Leper^2-\pi ^2 \left(2 \eper^2-5\right) (f_2 f_3+f_1 (9 f_2+4 f_3))\right) (5 e_y e_z-\j_y \j_z)\\
&\quad +\frac{1}{4 \eper} \epssa \left(\sqrt{1-\frac{1}{\eper^2}} \left(4 \eper^2-1\right)+3 \eper \Leper\right) (5 e_x e_z-\j_x \j_z); \\
\nonumber \j_z' &= -\frac{3}{\eper f_1 f_2 f_3} \sqrt{1-\frac{1}{\eper^2}} \epssa \Leper^3 \left(\left(8 \eper^2-5\right) (f_2 f_3+f_1 (f_2+f_3)) \Leper^2-\pi ^2 \left(2 \eper^2-5\right) (f_2 f_3+f_1 (9 f_2+4 f_3))\right) \left(5 e_x^2-5 e_y^2-\j_x^2+\j_y^2\right)\\
&\quad -\frac{1}{2 \eper} \sqrt{1-\frac{1}{\eper^2}} \left(\eper^2-1\right) \epssa (5 e_x e_y-\j_x \j_y).
\end{align}
\end{subequations}

\label{lastpage}

\end{document}